\documentclass[preprintnumbers,prd,twocolumn,showpacs,floatfix,preprintnumbers,superscriptaddress,nofootinbib]{revtex4-1}
\usepackage{graphicx}
\usepackage{epsfig}
\usepackage{bm}
\usepackage{amssymb}
\usepackage{float}
\usepackage{amsmath}
\usepackage{subfigure}
\usepackage{dcolumn}
\usepackage{cancel}
\usepackage[colorlinks]{hyperref}
\usepackage[usenames,dvipsnames]{color}
\hypersetup{
     breaklinks=true,
    pdfstartview={FitH},    
    colorlinks=true,       
    linkcolor=blue,          
    citecolor=red,        
    filecolor=magenta,      
    urlcolor=blue,           
    anchorcolor=green,      
    linktocpage=true
}

\newcommand{\Mpl}{M_{\textrm{Pl}}}
\renewcommand{\(}{\left(}
\renewcommand{\)}{\right)}

\newcommand{\nn}{\nonumber}
\def\al{\alpha}
\def\bet{\beta}
\def\gam{\gamma}

\def\Om{\Omega}
\def\sig{\sigma}
\def\Lam{\Lambda}

\def\ups{\upsilon}

\def\H{\mathcal{H}}
\def\F{\mathcal{F}}

\def\S{\mathcal{S}}
\def\P{\mathcal{P}}
\def\A{\mathcal{A}}
\def\B{\mathcal{B}}

\def\D{\mathcal{D}}

\def\K{\mathcal{K}}
\def\Q{\mathcal{Q}}

\def\del{\delta}
\def\ups{\upsilon}
\def\doi{http://doi.org}
\def\arxiv{http://arxiv.org/abs}
 \def\t{\tilde}
 \def\e{\mathrm{e}}
\def\r{\mathrm{r}}
\def\g{\mathrm{g}}

\def\m{\mathrm{m}}

\def\T{\mathrm{T}}
\def\s{\mathrm{s}}
\def\d{\mathrm{d}}

\def\vck{\vec k}

\allowdisplaybreaks

\begin{document}

\title{First and second order cosmological perturbations in light mass Galileon models}

\author{Md. Wali Hossain}
\email{wali.hossain@apctp.org}
\affiliation{Asia Pacific Center for Theoretical Physics, Pohang 37673, Korea}
\affiliation{Inter-University Centre for Astronomy and Astrophysics, Pune--411 007, India}

\begin{abstract}
In this paper, we investigate the first and second order cosmological perturbations in the light mass Galileon (LMG) scenario. LMG action includes cubic Galileon term along with the standard kinetic term and a potential which is added phenomenologically to achieve late time acceleration. The scalar field is nonminimally coupled to matter in the Einstein frame. Integral solutions of growing and decaying modes are obtained. The effect of the conformal coupling constant ($\bet$), at the perturbation level, has been studied. In this regard, we have studied linear power spectrum and bispectrum. Though different values of $\bet$ has different effects on power spectrum on reduced bispectrum the effect is not significant. It has been found that the redshift-space distortions (RSD) data can be very useful to constrain $\bet$. In this study we consider potentials which can lead to tracker behavior of the scalar field.
\end{abstract}

\pacs{98.80.-k, 95.36.+x, 04.50.Kd}

\graphicspath{{figs/}}

\maketitle

\section{Introduction}

Recent cosmological observations \cite{Riess:1998cb,Perlmutter:1998np,Ade:2015xua} unveil that our present Universe is expanding with an acceleration which is known as late time acceleration. At present, there is no proper theoretical explanation of it but one can have negative pressure, responsible for late time acceleration, from some exotic fluid known as {\it dark energy} \cite{Copeland:2006wr,Linder:2008pp,Silvestri:2009hh,Sahni:1999gb}. Cosmological constant ($\Lambda$) is the simplest candidate of dark energy and also consistent with all the cosmological observations. However, it is plagued with the fine tuning problem and cosmic coincidence problem. While the first problem is due to the small measured value compared to the theoretical value, the later problem asks the question why dark energy has become important only now?

Scalar fields can also behave like dark energy, known as quintessence field \cite{Wetterich:1987fk,Wetterich:1987fm,Ratra:1987rm}, with variable equation of state but mimicking cosmological constant during late time. Though this can not solve the fine tuning problem it can solve the cosmic coincidence problem for some specific scenarios known as tracker models \cite{Zlatev:1998tr,Steinhardt:1999nw}. In this scenario the scalar field energy density tracks the background energy density in the past and takes over matter during the recent past. The late time solution is an attractor solution for a wide range of initial condition. There are another class of models known as thawing models \cite{Scherrer:2007pu,Chiba:2009sj} in which the scalar field behaves as a cosmological constant in the past and starts evolving from recent past. Late time acceleration is transient in this scenario. 

It is also possible to explain late time acceleration without invoking any exotic component but modifying gravity known as modified theories of gravity (MOG) \cite{Fujii_Maeda,Clifton:2011jh,Hinterbichler:2011tt,Dvali:2000hr,Nicolis:2008in,deRham:2012az,deRham:2010kj,deRham:2014zqa,DeFelice:2010aj}. MOG generally possess an extra degree of freedom which can have effect in local physics. But local physics is extremely constraint from observation and well explained by Einstein's general theory of relativity. So one need to hide this extra degree of freedom to make the theory consistent for short distances too. One of the hiding or screening mechanisms is chameleon mechanism \cite{Khoury:2003aq,Khoury:2003rn}. It works for scalar fields which are nonminimally coupled to matter in Einstein frame and their effective mass depends on the local density. This mechanism can be implemented in many MOG like scalar-tensor theories \cite{Fujii_Maeda} and $F(R)$ theories \cite{DeFelice:2010aj}. Similar to chameleon mechanism symmetron mechanism \cite{Hinterbichler:2010es} also works for massive scalar fields but in symmetron mechanism, instead of the effective mass of the scalar field, the vacuum expectation value (VEV) of the scalar field depends on the local density. While the chameleon and symmetron mechanisms work for massive scalar fields for massless scalar fields the screening mechanism is known as Vainshtein mechanism \cite{Vainshtein:1972sx} proposed by A.I. Vainshtein in 1972 to circumvent the van
Dam-Veltman-Zakharov (vDVZ) discontinuity problem \cite{vanDam:1970vg,Zakharov:1970cc} in the linear theory massive gravity proposed by Pauli and Fierz in 1939 \cite{Fierz:1939ix}. Later this mechanism is implemented in many MOG {\it e.g.}, Dvali-Gabadadze-Porrati (DGP) theory \cite{Dvali:2000hr}, de Rham-Gabadadze-Tolley (dRGT) nonlinear massive gravity theory \cite{deRham:2010kj} and Galileon theories \cite{Nicolis:2008in}. 

Galileon is a scalar field appears in the decoupling limit of DGP action \cite{Luty:2003vm,Nicolis:2004qq}. In Minkowskian background the Galileon action respects the shift symmetry $\phi\to\phi+b_\mu x^\mu+c$, where $b_\mu$ and $c$ are constants. The shift symmetry makes the equation of motion of the Galileon field second order \cite{Nicolis:2008in} and free from Ostrogradsky
ghosts \cite{Woodard:2006nt} though the action contains higher derivative terms. Covariant form of the Galileon action was obtained in Ref.~\cite{Deffayet:2009wt} and it was shown that the equation of motion is still second order but the Galileon field is nonminimally coupled to curvature. Galileon theory can be a good alternative to dark energy which can produce late time acceleration \cite{Chow:2009fm,Silva:2009km,Kobayashi:2010wa,Kobayashi:2009wr,Gannouji:2010au,DeFelice:2010gb,DeFelice:2010pv,Ali:2010gr,Mota:2010bs,Deffayet:2010qz,deRham:2010tw,deRham:2011by,Heisenberg:2014kea}. Inflationary scenario, in the Galileon theory, has also been studied \cite{Creminelli:2010ba,Kobayashi:2010cm,Mizuno:2010ag,Burrage:2010cu,Kamada:2010qe}.

The decoupling limit of DGP action gives rise to cubic Galileon action of the form $(\partial_\mu\phi)^2\Box\phi$ \cite{Luty:2003vm,Nicolis:2004qq}. This term has some beautiful properties. On one hand, because of the shift symmetry, mentioned earlier, it gives second order equation of motion and on the other hand its nonlinear term is responsible for the Vainshtein effect which preserves the local physics by screening the Galileon field locally within a radius known as Vainshtein radius \cite{Nicolis:2008in}. Apart from this cubic Galileon term the Galileon Lagrangian has a linear term in $\phi$, standard kinetic term and two other higher derivative terms \cite{Nicolis:2008in,Deffayet:2009wt}. Galileon action can be realized as a particular form of Horndeski action \cite{Horndeski:1974wa}, the most general scalar tensor theory. It is shown in Ref.~\cite{Gannouji:2010au} that cubic Galileon theory (without linear potential term) can not give stable late time de Sitter solution without invoking at least one more higher derivative term. In order to get viable cosmology with the simplest Galileon correction {\it i.e.}, the cubic Galileon term, in Refs.~\cite{Ali:2012cv,Hossain:2012qm} a potential term has been considered phenomenologically which breaks the shift symmetry of the action but gives late time acceleration \cite{Ali:2012cv,Hossain:2012qm}. The nonlinear cubic Galileon term is still responsible for the Vainshtein mechanism \cite{Ali:2012cv}. Because of the potential and the requirements of late time cosmology the Galileon field will have a tiny mass and is dubbed as light mass Galileon. Background cosmology in LMG was studied for several potentials in Refs.~ \cite{Ali:2012cv,Hossain:2012qm}. 

Though the MOG can reproduce similar background evolution as $\Lam$CDM it can leave some distinguished features at the perturbation level. In this paper we are interested to carry out first order and second order cosmological perturbations in LMG. At the linear level of perturbation we shall study the linear growth and the power spectrum. We shall calculate the integral solutions of the growing and decaying modes by following the method depicted in Ref.~\cite{Bartolo:2013ws} (for other works on cosmological perturbation and structure formation in MOG see Refs.~\cite{Koyama:2009me,Kimura:2010di,Brax:2012sy,Barreira:2013eea,Li:2013tda,Wyman:2013jaa,Takushima:2013foa,Taruya:2014faa,Takushima:2015iha,Bellini:2015wfa}). In this work we generalize some results obtained in Ref.~\cite{Bartolo:2013ws} for nonminimal case with any potential. We would also compare the model with the redshift-space distortions (RSD) data \cite{Hudson:2012zga,Beutler:2012px,Howlett:2014opa,Percival:2004fs,Song:2008qt,Blake:2011rj,Samushia:2011cs,Tojeiro:2012rp,Gil-Marin:2015sqa,Tegmark:2006az,Blake:2012pj,Chuang:2013wga,Macaulay:2013swa,Guzzo:2008ac,delaTorre:2013rpa,Okumura:2015lvp}. RSD data can be very useful for constraining MOG \cite{Okada:2012mn,Tsujikawa:2012hv,Dodelson:2013sma,Nesseris:2014mea,DeFelice:2016ufg,Park:2016jym,Nersisyan:2017mgj}. At the level of second order perturbation we shall study the matter bispectrum in LMG. Though observations on cosmic microwave background \cite{Ade:2015lrj} indicate that there is no non-Gaussianity in the primordial fluctuations non-Gaussianity can be generated during late time matter fluctuations even though the initial fluctuations are Gaussian. This is because of the nonlinearities in the fluid equations. Linear theory of matter perturbation breaks for $k>0.1 h \rm Mpc^{-1}$ and nonlinear effects become important (for review on nonlinear cosmological perturbation see Ref.~\cite{Bernardeau:2001qr}). First higher order statistical quantity to measure nonlinear effects is the bispectrum. In this paper we are interested to see the evolution of the late time non-Gaussianity or matter bispectrum in LMG. 

The structure of this paper is as follows. In Sec.~\ref{sec:lmg} we describe the LMG scenario and in Sec.~\ref{sec:Cosmology} we study the background dynamics. In Sec.~\ref{sec:PT} we study cosmological perturbation. The analysis of linear and second order perturbation are shown in SubSecs.~\ref{sec:CPT} and \ref{sec:2p} respectively. The results in terms of power spectrum and reduced bispectrum are given in the Sec.~\ref{sec:power}. The comparison of the scenario under consideration with the RSD data is also shown in the same section. Finally we summarize our results in Sec.~\ref{sec:conc}.

\section{Light Mass Galileon}
\label{sec:lmg}

We shall consider the following action in the Einstein frame \cite{Ali:2012cv,Hossain:2012qm}
\begin{align}
\S=&\int \d^4x\sqrt{-\g}\Bigl [\frac{\Mpl^2}{2} R-\frac{1}{2}(\nabla \phi)^2\Bigl(1+\frac{\al}{M^3}\Box \phi\Bigr) - V(\phi) \Bigr]\nn\\   &+ \S\Bigl[\psi;\vartheta^2(\phi)\g_{\mu\nu}\Bigr] \, ,
\label{eq:action}
\end{align}
where $\Mpl=1/\sqrt{8\pi G}$ is the reduced Planck mass, $M$ is an energy scale, $\al$ is a dimensionless constant and $V$ is the potential for the field. $\vartheta(\phi)$ is the conformal factor which relates the Jordan frame metric ($g^{(\rm J)}_{\mu\nu}$) with the Einstein frame metric ($g_{\mu\nu}$) through the relation $g^{(\rm J)}_{\mu\nu}=\vartheta^2(\phi)g_{\mu\nu}$. For equivalence of this two frames, see Refs.~\cite{Deruelle:2010ht,Gong:2011qe,Chiba:2013mha}. Presence of the nonminimal coupling modifies the continuity equations of the different component of the Universe such a way so that the total energy density is still conserved. Action~(\ref{eq:action}) corresponds to the coupled quintessence \cite{Amendola:1999er} with a Galileon like correction $(\nabla \phi)^2 \Box \phi$. 

Variation of the action (\ref{eq:action}) with respect to (w.r.t.) the metric $g_{\mu\nu}$ gives Einstein equation  \begin{align}
\Mpl^2 G_{\mu\nu}= T_{(\m)\mu\nu}+T_{(\r)\mu\nu}+T_{(\phi)\mu\nu} \, ,
\label{eq:ee}
\end{align}
where
\begin{align}
T_{(\phi)\mu\nu}=& \phi_{;\mu}\phi_{;\nu}-\frac{1}{2}\g_{\mu\nu}(\nabla\phi)^2 -\g_{\mu\nu}V(\phi)\nn\\
&+\frac{\alpha}{M^3} \Bigl[\phi_{,\mu}\phi_{;\nu}\Box\phi+\g_{\mu\nu}\phi_{;\lambda}\phi^{;\lambda\rho}\phi_{;\rho} \nn \\ & - \phi^{;\rho}\(\phi_{;\mu}\phi_{;\nu \rho}+\phi_{;\nu}\phi_{;\mu \rho}\)\Bigr] \, ,
\label{eq:emt_phi}
\end{align}
and variation w.r.t. the scalar field $\phi$ gives the equation of motion of the LMG 
\begin{align}
& \Box \phi+\frac{\alpha}{M^3}\Bigl[(\Box\phi)^2-\phi_{;\mu\nu}\phi^{;\mu\nu}-R^{\mu\nu}\phi_{;\mu}\phi_{;\nu}\Bigr]-V'(\phi)\nn\\  &=\frac{\vartheta'(\phi)}{\vartheta(\phi)}T^{(\m)}=-\frac{\beta(\phi)}{\Mpl}T^{(\m)} \, ,
\label{eq:eom_phi}
\end{align}
where $'$ denotes the derivative w.r.t. $\phi$ and 
\begin{equation}
 \bet(\phi)=-\Mpl\frac{\vartheta'(\phi)}{\vartheta(\phi)}
\end{equation}
is the conformal coupling. Though we have started with a general form of $\bet$, later, for simplicity, we shall consider constant $\bet$ for which the conformal factor is exponential. Subscripts $m$, $r$ and $\phi$ in Eq.~(\ref{eq:ee}) and for the rest of the paper represent matter, radiation and the scalar field respectively.

\section{Background Cosmology}
\label{sec:Cosmology}

Let us consider the spatially flat Friedmann-Lema\^{i}tre-Robertson-Walker (FLRW) metric in conformal time $\tau$
\begin{equation}
\d s^2=a(\tau)^2\Big[-\d\tau^2+\d\vec{x}^2\Big] \, ,
\label{eq:metric}
\end{equation}
In which the Friedmann equations take the form
\begin{align}
\frac{3\Mpl^2\H^2}{a^2} &=\rho_\m+\rho_\r+\rho_\phi\,,
\label{eq:Fried1}\\
\frac{\Mpl^2}{a^2}(2\dot \H +\H^2)&=-\frac{\rho_\r}{3}-p_\phi\,,
\label{eq:Fried2}
\end{align}
and the equation of motion of the LMG is given by
\begin{align}
&\ddot{\phi}+2\H\dot{\phi}-\frac{3\alpha}{M^3a^2} \dot{\phi}\Bigl(2\H\ddot\phi+\dot\H\dot\phi\Bigr) +a^2 V'(\phi) \nn \\ & =-\frac{\beta(\phi)}{\Mpl}a^2 \rho_\m \, ,
\label{eq:eom_phi}
\end{align}
where $\H$ is the conformal Hubble parameter and a {\it dot} represents derivative w.r.t. the conformal time $\tau$, a convention which we shall follow for the rest of the paper. Hubble parameter tells us about the expansion history of the Universe. Using the Hubble parameter the effective equation of state (EoS) of the Universe can be written as
\begin{equation}
 w_{\rm eff}= -\frac{1}{3}\(1+2\frac{\dot\H}{\H^2}\)=\frac{p_\m+p_\r+p_\phi}{\rho_\m+\rho_\r+\rho_\phi} \, ,
 \label{eq:weff}
\end{equation}
$\rho$'s and $p$'s represent energy density and pressure respectively and 
\begin{eqnarray}
 \rho_\phi &=& \frac{\dot{\phi}^2}{2a^2}\Bigl(1-\frac{6\alpha}{M^3a^2}\H\dot\phi\Bigr)+V(\phi) \\
 p_\phi &=& \frac{\dot{\phi}^2}{2a^2}\Bigl(1+\frac{2\alpha}{M^3a^2}\(\ddot\phi-\H\dot\phi\)\Bigr)-V(\phi) \, .
\end{eqnarray}
Last two equations give us the EoS parameter for the scalar field
\begin{eqnarray}
 w_\phi=\frac{p_\phi}{\rho_\phi} \, .
 \label{eq:wphi}
\end{eqnarray}

As mentioned earlier the conservation equation for matter gets modified due to the presence of nonminimal coupling but for radiation there is no modifications as the scalar field couples with the trace of the energy momentum tensor of the component. So the conservation equations are given by
\begin{align}
\dot\rho_\m+3\H\rho_\m &=\frac{\beta(\phi)}{\Mpl}\dot{\phi} \rho_\m \, ,
\label{eq:cont_matt}\\
\dot\rho_\r+4\H\rho_\r &=0 \, ,\\
\dot\rho_\phi+3\H\rho_\phi (1+w_\phi) &=-\frac{\beta(\phi)}{\Mpl}\dot{\phi} \rho_\m \, .
\label{eq:cont_rad}
\end{align}
From the last three equations it is clear that the total energy density $\rho_{\rm tot}=\rho_\m+\rho_\r+\rho_\phi$ follows the standard evolution equation.


In this paper we consider potentials which can lead to tracker behavior of the scalar field. Not all potentials can give rise to tracker behavior. Potentials with steep region followed by a shallow region are suitable for tracker like dynamics. Considering this the following two potentials \cite{Barreiro:1999zs,Sahni:1999qe} are suitable for our analysis
\begin{equation}
V(\phi)=V_0\left[\e^{-\frac{\mu_1 \phi}{\Mpl}}+\e^{-\frac{\mu_2\phi}{\Mpl}}\right] \, ,
\label{eq:pot1}
\end{equation}
and
\begin{equation}
V(\phi)=V_0\left[\cosh\left(\frac{\zeta\phi}{\Mpl}\right)-1\right]^m \, ,
\label{eq:pot2}
\end{equation}
where $\mu_1$, $\mu_2$, $\zeta$ and $m$ are constants. $V_0$ fixes the energy scale of dark energy. These two potentials can give rise to tracker dynamics with late time acceleration. In potential~(\ref{eq:pot1}) the change of slope is responsible for the slow roll of the scalar field during late time when $\mu_1$ and $\mu_2$ have same sign. If there is a relative sign between these two parameters then the late time dynamics of the scalar field may be oscillatory. For potential~(\ref{eq:pot2}) late time dynamics is always oscillatory as long as $m>0$. Late time acceleration can be achieved for small ($<1$) values of $m$. Smaller the value of $m$ makes the dynamics closer to the de Sitter solution. In the LMG scenario,  potentials~\eqref{eq:pot1} and \eqref{eq:pot2} have been studied in Ref.~\cite{Ali:2012cv}. In this work, for numerical purpose, we shall consider the potential~\eqref{eq:pot1}.

\begin{figure}[h]
\centering
\includegraphics[scale=.75]{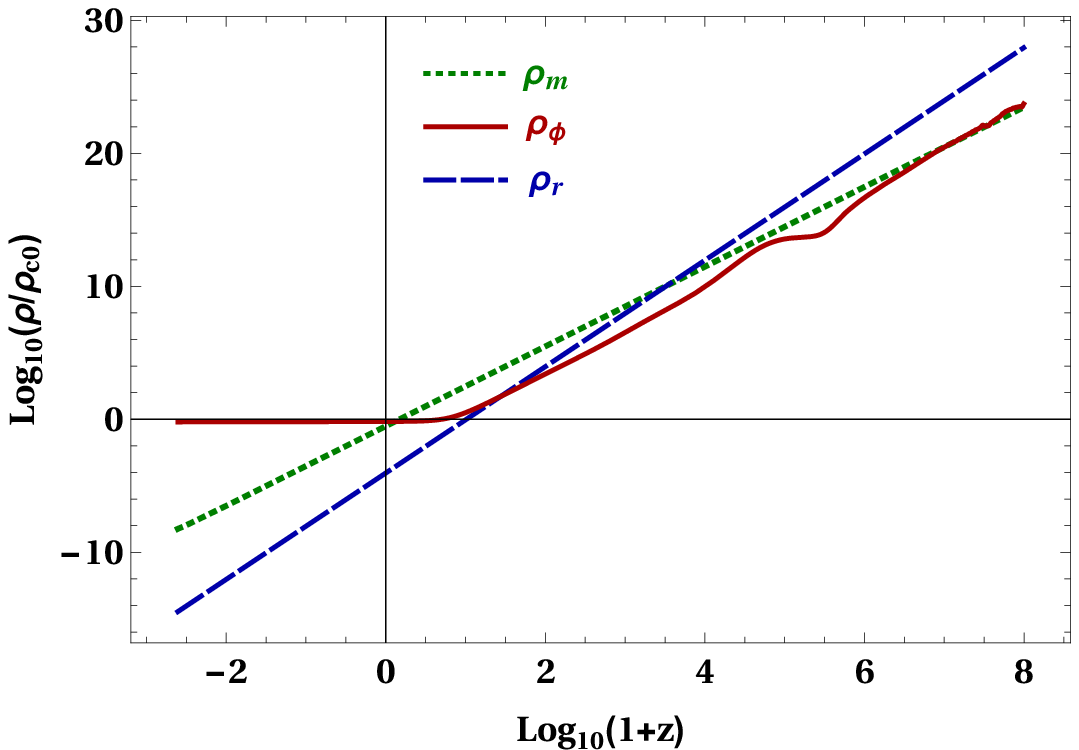}\vskip15pt
\includegraphics[scale=.75]{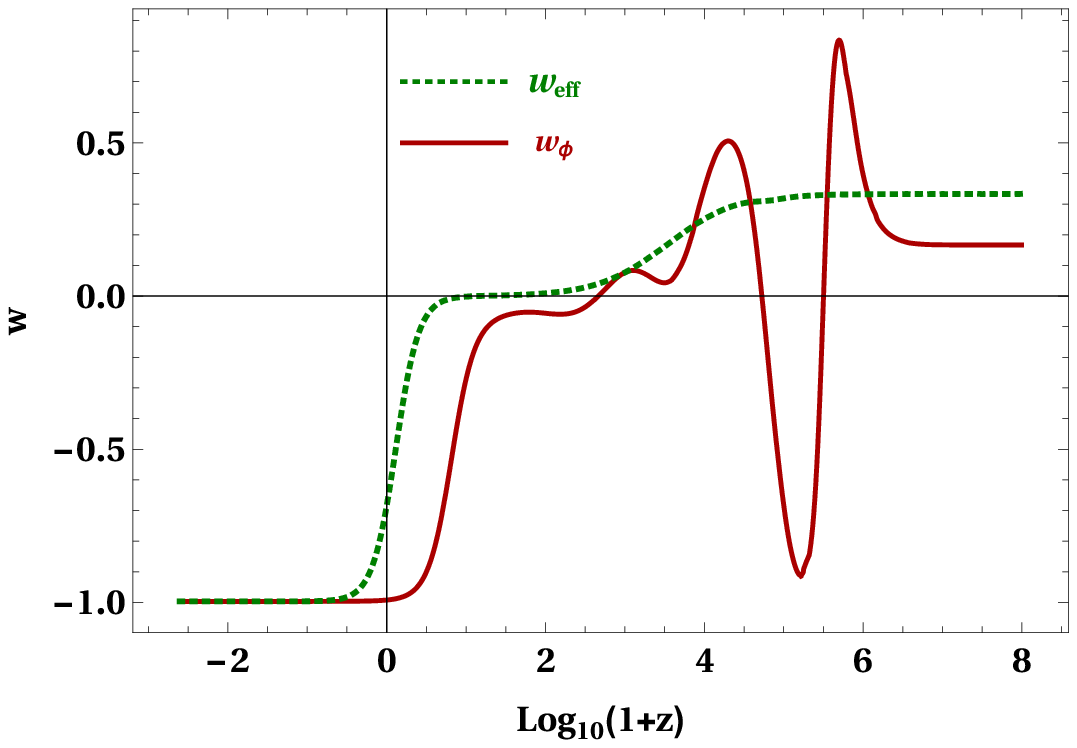}
\caption{{\bf Upper:} Green (dotted), blue (dashed) and red (solid) lines represent evolution of the energy density of matter, radiation and scalar field respectively. $\rho_{\rm c0}$ is the present critical density of the Universe. {\bf Lower:} Evolution of EoSs are shown. Red (solid) line represents scalar field EoS and green (dotted) line represents effective EoS. Both of the plots are for potential~(\ref{eq:pot1}) with $\mu_1=20$, $\mu_2=0.1$ and $\bet=0.01$.} 
\label{fig:back}
\end{figure}

In Fig.~\ref{fig:back} we show the background evolution in LMG. The upper figure of Fig.~\ref{fig:back} represents the evolution of the energy densities of different components of the Universe. It shows the tracker behavior of the scalar field. The scalar field tracks the background until recent past before it takes over matter and starts dominating. The Universe evolves through radiation and matter domination to scalar field domination giving rise to late time acceleration. The lower figure of Fig.~\ref{fig:back} shows the evolution of the effective EoS~(\ref{eq:weff}) and the scalar field EoS~(\ref{eq:wphi}). We can see that the scalar field EoS is close to $-1$ during recent time. The evolution of the whole Universe is represented by the effective EoS.

\section{Cosmological Perturbation}
\label{sec:PT}

In this section we shall discuss the linear and second order cosmological perturbations in LMG. We consider the following metric in the Newtonian gauge
\begin{equation}
\d s^2=a(\tau)^2\Big[-(1+2\Phi)d\tau^2+(1-2\Psi)\d\vec{x}^2\Big] \, ,
\label{eq:pmetric}
\end{equation}
where $\Phi$ and $\Psi$ are the scalar perturbations of the metric and are same as the gauge invariant Bardeen's potentials in this gauge \cite{Bardeen:1980kt}.

In this work, our aim is to calculate up to second order perturbation and we shall do order by order perturbation calculations. For this purpose one needs to expand any perturbation in series as shown below
\begin{eqnarray}
 \Phi &=& \Phi_1+\frac{1}{2!}\Phi_2+\frac{1}{3!}\Phi_3+\cdots \, , \\
 \Psi &=& \Psi_1+\frac{1}{2!}\Psi_2+\frac{1}{3!}\Psi_3+\cdots \, ,
\end{eqnarray}
where the subscripts 1, 2, 3 etc. represent the order of perturbation. To see the matter bispectrum we shall consider up to second order perturbation.

The fluid four velocity $u^\mu$ satisfies the relation $u^\mu u_\mu=-1$, which gives us the components of the velocity perturbation $\del u^\mu$ (up to second order)
\begin{eqnarray}
 \del u^0 &=& \frac{1}{a(\tau)}\(-\Phi_1-\frac{1}{2}\Phi_2+\frac{3}{2}\Phi_1^2+\frac{1}{2}\ups_{1i}\ups_1^i\) \, ,\\
 \del u^i &=& \frac{1}{a(\tau)}\(\ups_1^i+\frac{1}{2}\ups_2^i\) \, ,
\end{eqnarray}
where $\ups^i$ is the spatial three velocity of the fluid. 

In the following subsections we discuss linear and second order cosmological perturbations in LMG. First we perform the linear perturbation theory and calculate the integral solutions for the growing and decaying modes. Then using those solutions and doing second order perturbation we calculate the matter bispectrum.

\subsection{Linear perturbation}
\label{sec:CPT}

Linear perturbation in the energy momentum tensor of the LMG field gives
\begin{align}
 \delta T^{(\phi)0}_{(1)~0} &=-\frac{1}{a^2}\bigg[\dot{\phi}\(1-\frac{9\al}{M^3a^2}\H\dot{\phi}\)\dot{\del\phi_1}+
a^2V'(\phi)\del\phi_1   \nn \\ & +\frac{\dot{\phi}^2\al}{a^2M^3}\nabla^2\del\phi_1-\dot{\phi}^2\(1-\frac{12\al}{M^3a^2}\H\dot\phi\)\Phi_1 \nn \\ & +\frac{3\al}{M^3 a^2}\dot{\phi}^3\dot\Psi_1\bigg] 
\label{eq:tp00} \\
 \delta T^{(\phi)0}_{(1)~i} &=\frac{\dot{\phi}}{a^2}\partial_i\bigg[\frac{\al}{M^3a^2}\dot{\phi}\dot{\del\phi_1}+\del\phi_1\left(1-\frac{3\al}{M^3a^2}\H\dot{\phi} \right) \nn \\ & -\frac{\al}{M^3a^2}\dot{\phi}^2\Phi_1\bigg]  =-\del T^{(\phi)i}_{(1)~0} \, 
 \label{eq:tpi0} \\
 \delta T^{(\phi)i}_{(1)~j} &=\frac{1}{a^2}\bigg[\frac{\al}{M^3a^2}\dot{\phi}^2\ddot{\del\phi_1}+\dot{\phi}\dot{\del\phi_1}\(1+\frac{\al}{M^3a^2}\(2\ddot{\phi}-3\H\dot\phi\)\) \nn \\ & -a^2 V'(\phi) \del\phi_1  -\dot{\phi}^2\Phi_1 \(1+\frac{4\al}{M^3a^2}\(\ddot{\phi}-\H\dot\phi\)\) \nn \\ & - \frac{\al}{M^3a^2}\dot{\phi}^3\dot{\Phi}_1\bigg]\delta^i_j \, ,
\label{eq:tpij}
\end{align}
where $\del T^{(\phi)}$'s are the different perturbed components of the energy momentum tensor of the LMG and $\del\phi$ is the perturbation in the field $\phi$.

Let us introduce the density contrast $\del$ which is defined as 
\begin{equation}
 \del=\frac{\rho_\m}{\bar\rho_\m}-1 \, .
 \label{eq:density_cont}
\end{equation}
$\del$ can also be expanded as $\del=\del_1+(1/2)\del_2+\dots$. From now on, a $bar$ above $\rho$'s and $p$'s denotes the corresponding unperturbed quantities.

The perturbed equation of motion of the field is given by
\begin{eqnarray}\nonumber
&& \left(1-\frac{6\alpha}{M^3a^2}\H\dot{\phi}\right)\ddot{\del\phi_1}+2\left(\H- \frac{3\al}{M^3a^2}\(\H\ddot{\phi}+\dot{\phi}\dot{\H}\)\right)\dot{\delta\phi_1} \nn \\* && + a^2 V''(\phi)\del\phi_1 - \left(1-\frac{2\al}{M^3a^2}(\ddot{\phi}+\H\dot{\phi}) \right) \nabla^2\del\phi_1 \nn \\ && - \dot\phi\left(1-\frac{9\al}{M^3a^2}\H\dot\phi\right)\dot\Phi_1 +
2\bigg[\(V'(\phi)+\frac{\bet}{\Mpl}\rho_\m\)a^2 \nn \\  &&- \frac{3\al\dot\phi}{M^3a^2}\left(2\H\ddot{\phi}+\dot{\phi}\dot\H\right)\bigg] \Phi_1+ \frac{\al}{M^3a^2}\dot\phi^2\nabla^2\Phi_1 \nn \\  &&+ \frac{3\al}{M^3a^2}\dot\phi^2\ddot\Psi_1  -3\dot\phi\left(1-\frac{3\al}{M^3a^2}(2\ddot{\phi}+ 3\H\dot{\phi})\right)\dot\Psi_1 \nn \\  && =-\frac{\bet}{\Mpl}a^2\del_1\bar\rho_\m-\frac{\bet'(\phi)}{\Mpl}a^2\bar\rho_\m\del\phi_1
\label{pieom}
\end{eqnarray}

From the off diagonal part of the $ij$ components of the perturbed Einstein's tensor (see Appendix~\ref{app_sub}) and the energy momentum tensor we can have 
\begin{equation}
 \Phi_1=\Psi_1 \, .
\end{equation}
So at the linear level, in the scenario under consideration, there is no gravitational slip.

From the continuity equation and at the linear level we obtain
\begin{align}
&\dot{\del_1}-3\dot{\Psi}_1-\nabla^2\ups_1 =\frac{\beta(\phi)}{\Mpl}\dot{\del\phi_1}+\frac{\bet'(\phi)}{\Mpl}\dot\phi \del\phi_1 \, ,
\label{eq:deltamdot}  \\
&\dot{\ups}_1+\(\H+\frac{\bet(\phi)}{\Mpl}\dot{\phi}\)\ups_1-\Phi_1 =\frac{\bet(\phi)}{\Mpl}\del\phi_1 \, .
\label{eq:vdot}
\end{align}
Here we have introduced the velocity potential $\ups$ which is defined as $\ups^i=-\partial_i\ups$ and $\ups=\ups_1+(1/2)\ups_2$ (for linear perturbation $\ups_2=0$).

In subhorizon ($k^2\gg\H^2$) and quasistatic ($|\ddot\phi|\lesssim \H|\dot\phi|\ll k^2|\phi|$) approximations, from Eqs.~(\ref{eq:deltamdot}) and (\ref{eq:vdot}) we get \cite{Ali:2012cv} (perturbed quantities in subhorizon and quasistatic approximations for linear perturbation are given in Appendix~\ref{app_sub})
\begin{eqnarray}
 \ddot\del_1+\(\H+\frac{\bet(\phi)}{\Mpl}\dot\phi\)\dot\del_1-4\pi G_{\rm eff}a^2\bar\rho_\m\del_1=0 \, ,
 \label{eq:den_con_evo}
\end{eqnarray}
where
\begin{eqnarray}
 G_{\rm eff} &=& G\(1+\frac{\left(Q+2\bet(\phi)\)^2}{2P-Q^2}\) \, , \\
  P &=& 1-\frac{2\al}{M^3a^2}\(\ddot\phi+\H\dot\phi\) \, , \\
 Q &=& \frac{\al}{M^3a^2\Mpl}\dot\phi^2 \, .
\end{eqnarray}
Eq.~(\ref{eq:den_con_evo}) defers from the standard form (for $\Lam$CDM) of the evolution equation for the density contrast in the second and third terms of the left hand side and can be regained by putting $\bet(\phi)=\al=0$. Modification of gravity is encoded in the effective Newton's constant $G_{\rm eff}$. Here one should note that the modification of gravity can also be seen in the evolution of the Hubble parameter and that can give a different growth history of cosmological perturbation even if the the Newton's constant is not modified. 

The solution of Eq.~(\ref{eq:den_con_evo}) can be written as the linear superposition of two independent solutions
\begin{eqnarray}
 \del_1(\tau,\vec k)=c_+ D_+(\tau)\del_1(\vec k,0)+c_- D_-(\tau)\del_1(\vec k,0) \, ,
 \label{eq:del_sol}
\end{eqnarray}
where $c_+$ and $c_-$ are the constants, $D_+$ and $D_-$ are the growing and decaying modes respectively and $\del_1(\vec k,0)$ is the primordial density fluctuation. 

\begin{figure}[h]
\centering
\includegraphics[scale=.9]{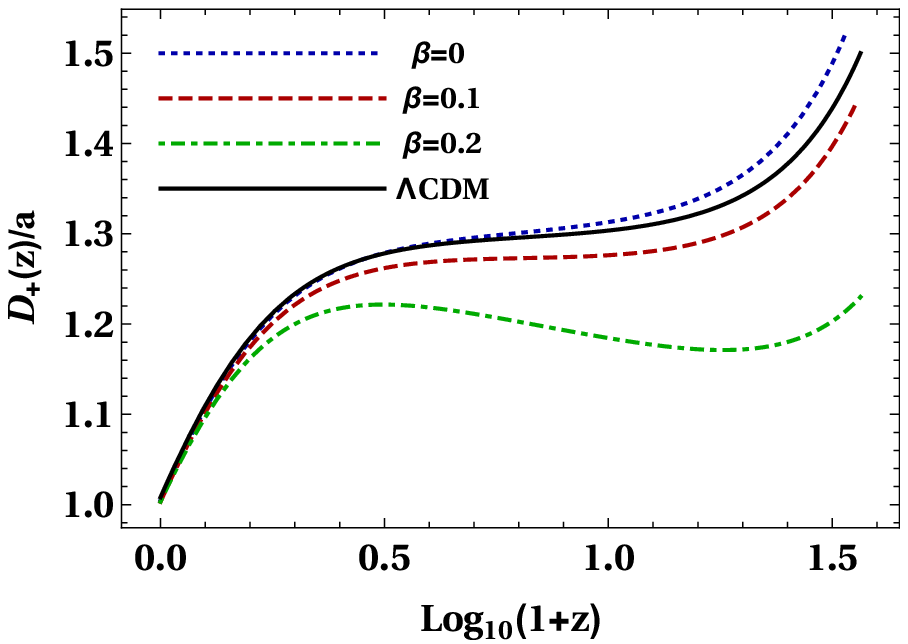}\vskip15pt
\includegraphics[scale=.9]{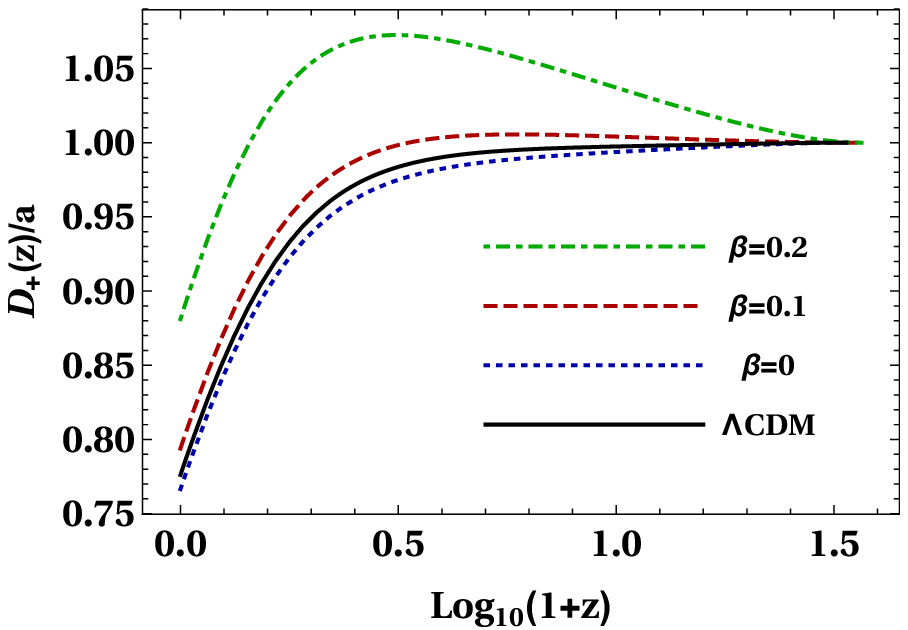}
\caption{Evolution of the growing mode $D_+(z)$ is shown for two different normalization for the potential~\eqref{eq:pot1} with $\mu_1=20$, $\mu_2=0.1$. In the upper figure we have normalized $D_+(z)$ at present while in the lower figure it is normalized in the past. In both the figures blue (dotted), red (dashed), green (dotdashed) and black (solid) lines represent curves corresponding to $\bet=0,\; 0.1,\; 0.2$ and $\Lam$CDM respectively. $D_+(z)/a=1$ for E-dS Universe. } 
\label{fig:den_con12}
\end{figure}

In Fig.~\ref{fig:den_con12} we have shown the evolution of the growing mode ($D_+(z)$) by solving Eq.~\eqref{eq:den_con_evo} numerically. Both the figures are normalized to Einstein-de Sitter (E-dS) Universe but at different times. While the upper figure of Fig.~\ref{fig:den_con12} is normalized at present the lower figure of Fig.~\ref{fig:den_con12} is normalized in the past. We can see that while we change the normalization the relative positions of different curves for different $\bet$ change. This same change we can also observe in the power spectrum for two different normalizations.

Next we shall find out the integral solutions for the growing and decaying modes for the scenario~(\ref{eq:action}).
\subsubsection{Growing and decaying modes: Integral solutions}
\label{sec:gro_dec}

To calculate the growing and decaying modes we need to specify the form of $\bet(\phi)$ and for simplicity, in this work, we consider a constant $\bet(\phi)$ {\it i.e.}, $\bet(\phi)=\rm constant=\bet$. Now let us consider the following transformation
\begin{equation}
 \t a=a~\e^{\bet\phi/\Mpl} \, ,
 \label{eq:atilde}
\end{equation}
which rewrites Eq.~(\ref{eq:den_con_evo}) as
\begin{equation}
  \ddot\del_1+\t\H\dot\del_1-4\pi \t G_{\rm eff}\t a^2\bar\rho_\m\del_1=0 \, ,
 \label{eq:den_con_evo1}
\end{equation}
where
\begin{eqnarray}
 \t\H &=& \frac{1}{\t a}\frac{\d \t a}{\d \tau}= \H+\frac{\bet}{\Mpl}\dot\phi=\H\frac{\d \ln\t a}{\d\ln a} \, . 
 \label{eq:Htilde}\\
 \t G_{\rm eff} &=& G_{\rm eff} ~\e^{-(2\bet/\Mpl)\phi} \, .
 \label{eq:Gefftilde}
\end{eqnarray}

Eq.~(\ref{eq:den_con_evo1}) has the same form as the standard evolution equation of the density contrast with the new scale factor ($\t a$), Hubble parameter ($\t \H$) and effective Newton's constant ($\t G_{\rm eff}$). 

To solve the Eq.~(\ref{eq:den_con_evo1}) we shall follow the procedure depicted in Ref.~\cite{Bartolo:2013ws}. Using $\t a$ as a new time variable Eq.~(\ref{eq:den_con_evo1}) can be written as
\begin{eqnarray}
 \frac{\d^2\del_1(\t a)}{\d \t a^2}+\(\frac{2}{\t a}+\frac{1}{\t\H}\frac{\d\t\H}{\d \t a}\)\frac{\d\del_1(\t a)}{\d \t a}-A(\t a)\del_1(\t a)=0 \, ,~~~
 \label{eq:den_con_evo12}
\end{eqnarray}
where
\begin{eqnarray}
 A(\t a)=4\pi \t G_{\rm eff}\frac{\bar\rho_\m}{\t\H^2}\, .
\end{eqnarray}

\begin{figure}[h]
\centering
\includegraphics[scale=.9]{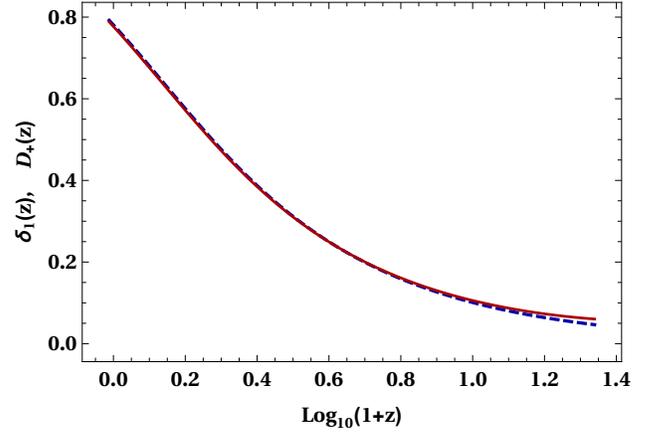}
\caption{Red (solid) line represents the numerical evolution of the growing mode~\eqref{eq:Dp} and the dashed (blue) line is the solution of the Eq.~\eqref{eq:den_con_evo} for the potential~\eqref{eq:pot1} with $\mu_1=20$, $\mu_2=0.1$ and $\bet=0.1$.} 
\label{fig:den_con}
\end{figure}

Solution of the Eq.~(\ref{eq:den_con_evo12}) is given in the Appendix~\ref{app_int}. The growing mode, $D_+(\t a)$, can be related to $D_1(\t a)$ (Eq.~(\ref{eq:D1})) and decaying mode, $D_-(\t a)$, can be related to the combination of $D_1(\t a)$ and $D_2(\t a)$ (Eq.~(\ref{eq:D1}) and (\ref{eq:D2})) and are given by
\begin{eqnarray}
 D_+(\t a)&=&\t a_\m^{7/4}\e^{-3\bet\phi_\m/4\Mpl}\(\frac{\A(\phi_\m)}{\A(\phi_0)}\)^{1/2}D_1(\t a)  \nn \\* 
 &=&\t a_\m^{7/4}\e^{-3\bet\phi_\m/4\Mpl}\(\frac{\A(\phi_\m)}{\A(\phi_0)}\)^{1/2} \frac{\gam(\t a)}{\t a}\sqrt{\frac{\t\H_0}{\t\H}} , ~~~~~ 
 \label{eq:Dp}\\
 D_-(\t a)&=&\t a_\m^{-3/4}\e^{7\bet\phi_\m/4\Mpl}\(\frac{\A(\phi_\m)}{\A(\phi_0)}\)^{1/2} \frac{\gam(\t a)}{\t a}\sqrt{\frac{\t\H_0}{\t\H}} \nn \\
 && \times \bigg[1-\frac{5}{2}\frac{1}{\t a_\m\A(\phi_\m)}\int_{\t a_\m}^{\t a}\frac{\d \t a'}{\gam^2(\t a')}\bigg] \, .
 \label{eq:Dm}
\end{eqnarray}
where $\gam(\t a)$ and $\A(\phi)$ are defined in the Eqs.~(\ref{eq:gam_ai}) and (\ref{eq:Aphi}) respectively. Fig.~\ref{fig:den_con} compares the evolution of the integral solution of the growing mode \eqref{eq:Dp} with the one obtained from the evolution equation of the density contrast~\eqref{eq:den_con_evo}. Red (solid) line represents Eq.~\eqref{eq:Dp} and blue (dashed) line represents the solution of the Eq.~\eqref{eq:den_con_evo} and it is clear that the integral solution of the growing mode has the same evolution as the solution of Eq.~\eqref{eq:den_con_evo}.

\subsection{Second order perturbation}
\label{sec:2p}

In this subsection we proceed to the second order perturbation which allows us to analyze the scenario in a mildly nonlinear regime. In second order perturbation we get two kind of terms, one is linear in the second order perturbations ({\it e.g.}, $\Psi_2$, $\Phi_2$ etc.) and the second one is quadratic terms of the linear perturbations ({\it e.g.}, $\Psi_1^2$, $\Psi_1\Phi_1$ etc.). Structures of the perturbation equations are same as the linear case except for the quadratic terms of the first order perturbations. 

Second order perturbation of the matter continuity equation gives
\begin{eqnarray}
 \dot{\del_2}-3\dot{\Psi}_2-\nabla^2\ups_2-\frac{\bet(\phi)}{\Mpl}\dot{\del\phi_2}-\frac{\bet'(\phi)}{\Mpl}\dot\phi \del\phi_2 &=& S_1 ,~
\label{eq:deltamdot2}  \\
\partial_i\(\dot{\ups}_2+\(\H+\frac{\bet(\phi)}{\Mpl}\dot{\phi}\)\ups_2-\Phi_2 -\frac{\bet(\phi)}{\Mpl}\del\phi_2\) &=& S_2 ,~~~~~
\label{eq:vdot2}
\end{eqnarray}
where
\begin{eqnarray}
 S_1 &=& 2(3\dot\Psi_1+\nabla^2\ups_1)\del_1+2\nabla^2\ups_1\Phi_1\nn \\ && -2\(\H+\frac{\bet(\phi)}{\Mpl}\dot\phi\)\partial_i\ups_1\partial^i\ups_1  -4\partial_i\dot\ups_1\partial^i\ups_1 \nn \\ && +2\partial_i\del_1\partial^i\ups_1  +4\partial_i\ups_1\partial^i\Phi_1-6\partial_i\ups_1\partial^i\Psi_1  +12\Psi_1\dot\Psi_1 \nn \\ && +2\frac{\bet(\phi)}{\Mpl} \del_1\dot{\del\phi_1}+2\frac{\bet'(\phi)}{\Mpl}\(\del\phi_1\dot{\del\phi_1}+\dot\phi\del_1\del\phi_1\) \nn \\* && +\frac{\bet''(\phi)}{\Mpl}\dot\phi \del\phi_1^2 \, , \\ 
 S_2 &=& 2\(-\dot\del_1+5\dot\Psi_1\)\partial_i\ups_1 +2\partial_j\(\partial_i\ups_1\partial^j\ups_1\)\nn \\ && -2\(\del_1-\Phi_1-2\Psi_1\)\partial_i\Bigg(\(\H+\frac{\bet(\phi)}{\Mpl}\dot\phi\)\ups_1+\dot\ups_1\Bigg) \nn \\ &&  +2(\del_1-2\Phi_1)\partial_i\Phi_1 +2\frac{\bet(\phi)}{\Mpl}\del_1\partial_i\del\phi_1 \nn \\ && +2\frac{\bet'(\phi)}{\Mpl}\del\phi_1\partial_i\del\phi_1 \, .
\end{eqnarray}

The evolution equation of the second order density contrast has the following form 
\begin{eqnarray}
 \ddot\del_2+\(\H+\frac{\bet}{\Mpl}\dot\phi\)\dot\del_2-4\pi G_{\rm eff}a^2\bar\rho_\m\del_2=S_\del \, ,
 \label{eq:den_con_evo2}
\end{eqnarray}
where $S_\del$ contains the terms quadratic in the first order perturbations and given by
\begin{eqnarray}
 S_\del &=& \dot S_1+\(\H+\frac{\bet}{\Mpl}\dot\phi\)S_1+\partial_i S_2 \nn \\* && -\frac{Q+2\bet}{\Mpl\(2P-Q^2\)}S_{\rm eom}-\frac{\(P+\bet Q\)a^2}{2P-Q^2}S_{\del1} ~ , ~~  
 \label{eq:S_del}
\end{eqnarray}
where
\begin{eqnarray}
 S_{\del1} &=& \frac{1}{\Mpl^2}S_{\rm T00}+S_{\rm G00}-\frac{1}{\Mpl^2}S_{\rm Tii}-S_{\rm Gii} \nn \\* && -\frac{4}{\Mpl^2}(\bar\rho_\m+\bar\rho_\phi+\bar p_\phi)\partial_i\ups_1\partial^i\ups_1 \, .
 \label{eq:sdel1}
\end{eqnarray}
$S_{\rm G00}$, $S_{\rm Gii}$ ($=S_{\rm Gij}$ with $i=j$), $S_{\rm T00}$, $S_{\rm Tii}$ ($=S_{\rm Tij}$ with $i=j$) and $S_{\rm eom}$ are defined in the Eq.~\eqref{eq:sg00}, \eqref{eq:sgij}, \eqref{eq:st00}, \eqref{eq:stij} and \eqref{eq:seom} in the Appendix~\ref{app_2nd} where the second order perturbation terms are written along with the subhorizon approximation.

Fourier transform of the source term (\ref{eq:S_del}) gives
\begin{eqnarray}
 \hat S_{\del}(\tau,\vck) &=& \int \d^3k_1\d^3k_2\del^{(3)}(\vck-\vck_1-\vck_2) \K(\tau,\vck_1,\vck_2) \nn \\* && \times\del_1(\tau,\vck_1)\del_1(\tau,\vck_2) \, ,
\end{eqnarray}
where $\del^{(3)}(\dots)$  is the three dimensional Dirac delta function and $\K(\tau,\vck_1,\vck_2)$ is the symmetrized kernel. To calculate the expression of $\K(\tau,\vck_1,\vck_2)$ we shall consider constant $\bet$. Now using the results of linear perturbation in subhorizon approximation the symmetrized kernel $\K(\tau,\vck_1,\vck_2)$ is given by
\begin{widetext}
 \begin{eqnarray}
  \K(\tau,\vck_1,\vck_2) &=& \Bigg\{2\Xi^2+\(1+\frac{(Q+2\bet)^2}{2P-Q^2}\)\frac{\bar\rho_\m a^2}{\Mpl^2}-\frac{2\al}{M^3}\(\frac{Q+2\bet}{2P-Q^2}\)^3\frac{\bar\rho_\m^2 a^2}{\Mpl^3}\Bigg\}
  +\Bigg\{2\Xi^2+\frac{1}{2}\(1+\frac{(Q+2\bet)^2}{2P-Q^2}\)\frac{\bar\rho_\m a^2}{\Mpl^2}\Bigg\} \nn \\ && \times \frac{\vck_1\cdot\vck_2(k_1^2+k_2^2)}{k_1^2k_2^2}
  +\Bigg\{2\Xi^2+\frac{2\al}{M^3}\(\frac{Q+2\bet}{2P-Q^2}\)^3\frac{\bar\rho_\m^2 a^2}{\Mpl^3}\Bigg\}\frac{(\vck_1\cdot\vck_2)^2}{k_1^2k_2^2}
  +\Bigg\{2(\dot{\t\H}+\t\H^2)\Xi^2+4\t\H(2r+F)\Xi \nn \\ && +4s\Xi+4rF+8\Xi G+4F^2-\frac{\al}{M^3a^2\Mpl}\frac{Q+2\bet}{2P-Q^2}\bigg(8\H W N-4N^2+8\dot\phi FN-8\dot\phi^2F^2+2\dot\H W^2 \nn \\ && -8\dot\phi GW-4(\ddot\phi+2\H\dot\phi)FW \bigg)-\frac{P+\bet Q}{2P-Q^2}\bigg(8F^2-\frac{2\al}{M^3a^2\Mpl^2}(\ddot\phi-2\H\dot\phi)W^2-\frac{8\al}{M^3a^2\Mpl^2}\dot\phi W N \nn \\ && +\frac{4}{\Mpl^2}(\bar\rho_\m+\bar\rho_\phi+\bar p_\phi)\Xi^2\bigg)\Bigg\}\frac{\vck_1\cdot\vck_2}{k_1^2k_2^2}
   +\Bigg\{3J+9\Xi G+4rF+\frac{\bet}{\Mpl}(L+\Xi N)+\t\H\(3G+4\Xi F+\frac{\bet}{\Mpl}N\) \nn \\ && +2F^2-\frac{2\al}{M^3a^2\Mpl^2}\frac{P+\bet Q}{2P-Q^2}\dot\phi WN+\frac{2}{\Mpl}\frac{Q+2\bet}{2P-Q^2}\Bigg(\(1-\frac{4\al}{M^3a^2}(\ddot\phi+\H\dot\phi)\)FW \nn \\ && -\frac{\al}{M^3a^2}\bigg(F^2-(L+\H N)W-\dot\phi NF+3\dot\phi GW\bigg)\Bigg)\Bigg\} \(\frac{1}{k_1^2}+\frac{1}{k_2^2}\)
   +\Bigg\{12(G^2+FJ)+12\t\H FG \nn \\ && -\frac{Q+2\bet}{2P-Q^2}\frac{1}{\Mpl}\Bigg(4\(1-\frac{12\al}{M^3a^2}\H\dot\phi\)FL+\(8-\frac{12\al}{M^3a^2}(\ddot\phi+6\H\dot\phi)\)GN-a^2V'''(\phi)W^2 \nn \\ && +8\(\H-\frac{6\al}{M^3a^2}(\H\ddot\phi+\dot\H\dot\phi)\)FN-4\dot\phi\(2-\frac{6\al}{M^3a^2}(\ddot\phi+6\H\dot\phi)\)FG+8\(\(V'(\phi)+\frac{\bet}{\Mpl}\bar\rho_\m\)a^2 \nn \right. \\ && \left. +\frac{6\al}{M^3a^2}\dot\phi(2\H\ddot\phi+\dot\H\dot\phi)\)F^2+\frac{\al}{M^3a^2}\bigg(12\dot\phi^2(FJ+2G^2)+12\H NL+6\dot\H N^2-12\dot\phi GL-12\dot\phi NJ\bigg)\Bigg) \nn \\ && +\frac{P+\bet Q}{2P-Q^2}\Bigg(48(\dot\H F+\H G)F+\frac{1}{\Mpl^2}\bigg(\(1-\frac{18\al}{M^3a^2}\H\dot\phi\)\(N^2-4\dot\phi(NF-\dot\phi F^2)\)+a^2V''(\phi)W^2 \nn \\ && +\frac{12\al}{M^3a^2}\dot\phi(N-2\dot\phi F)L+3\(1+\frac{2\al}{M^3a^2}(\ddot\phi-3\H\dot\phi)\)N^2 -4\dot\phi\(1+\frac{2\al}{M^3a^2}(2\ddot\phi-3\H\dot\phi)\)FN \nn \\ &&  +4\dot\phi^2\(1+\frac{6\al}{M^3a^2}(\ddot\phi-\H\dot\phi)\)F^2-a^2V''(\phi)W^2\bigg)\Bigg)\Bigg\}\frac{1}{k_1^2k_2^2}
   \label{eq:K}
 \end{eqnarray}
\end{widetext}
The functions $\Xi$, $r$ and $s$ relate first, second and third derivatives of $\del_1$ with $\del_1$ respectively. Similarly the functions $F$, $G$ and $J$ relate $\Phi_1$, $\dot\Phi_1$ and $\ddot\Phi_1$ (or $\Psi_1$ and its derivatives) with $\del_1$ respectively and functions $W$, $N$ and $L$ relate $\del\phi_1$, $\dot{\del\phi_1}$ and $\ddot{\del\phi_1}$ with $\del_1$. Explicit form of the above mentioned functions are given below
\begin{eqnarray}
  \Xi &=& \frac{\dot\gam}{\gam}-\t\H-\frac{1}{2}\frac{\dot{\t\H}}{\t\H} \, ,\\ 
  r &=& \dot \Xi+\Xi^2\, , \\
  s &=& \dot r+r\Xi \, .
\end{eqnarray}

\begin{eqnarray}
  F &=& -\frac{P+\bet Q}{2P-Q^2}\frac{\bar\rho_\m a^2}{\Mpl^2} \, , \\
  G &=& \dot F+\Xi F \, ,\\
  J &=& \dot G+\Xi G \, .
\end{eqnarray}

\begin{eqnarray}
  W &=& -\frac{Q+2\bet}{2P-Q^2}\frac{\bar\rho_\m a^2}{\Mpl} \, , \\
  N &=& \dot W+\Xi W \, ,\\*
  L &=& \dot N+\Xi N \, .
\end{eqnarray}

Homogeneous part of the inhomogeneous linear second order differential equation (ILDE)~(\ref{eq:den_con_evo2}) is similar as the Eq.~(\ref{eq:den_con_evo}). So the solution of the homogeneous part of the ILDE~(\ref{eq:den_con_evo2}) will have the same form as that of the solutions of Eq.~(\ref{eq:den_con_evo}). The general solution of the ILDE~(\ref{eq:den_con_evo2}) can be written by calculating the Wronskian and is given by

\begin{eqnarray}
 \del_2(\t a,\vck) &=& D_+(\t a)\del_2(\vck)-D_+(\t a)\int_{\t a_\m}^{\t a}\frac{D_-(\t a')\hat S_\del(\t a',\vck)}{\t a'^2\t\H^2(\t a')W_\r(\t a')}\d \t a' \nn \\* && +D_-(\t a)\int_{\t a_\m}^{\t a}\frac{D_+(\t a')\hat S_\del(\t a',\vck)}{\t a'^2\t\H^2(\t a')W_\r(\t a')} \d \t a'\, ,
 \label{eq:den2}
\end{eqnarray}
where $\del_2(\vck)$ is the initial second order matter perturbation and $W_\r$ is the Wronskian which is given by
\begin{eqnarray}
 W_\r (\t a) &=& D_+(\t a) \frac{\d D_-(\t a)}{\d \t a}- D_-(\t a) \frac{\d D_+(\t a)}{\d\t a}\nn \\ &=& -\frac{5}{2}\frac{\H_0}{\t a^2 \t \H}\e^{\bet\phi_\m/\Mpl}\, .
\end{eqnarray}

So the density contrast is given by
\begin{eqnarray}
 \del(\t a,\vck) &=& \del_1(\t a,\vck)+\frac{1}{2}\del_2(\t a,\vck) \nn \\* 
  &=& D_+(\t a) \del_1(\vck)+\int \d^3k_1\d^3k_2 \del^{(3)}(\vck-\vck_1-\vck_2)\nn \\* && \times \F_2(\t a,\vck_1,\vck_2) \del_1(\t a,\vck_1)\del_1(\t a,\vck_2) \, ,
\end{eqnarray}
where 
\begin{eqnarray}
 \F_2(\t a,\vck_1,\vck_2) &=& \int_{\t a_\m}^{\t a}\d\t a' ~\frac{\D(\t a,\t a')\K(\t a',\vck_1,\vck_2)}{2\t a'^2\t\H^2(\t a')W_\r(\t a')} \,,\\
 \D(\t a,\t a') &=& \frac{D_+^2(\t a')}{D_+^2(\t a)} \Big(D_-(\t a)D_+(\t a') \nn \\* && -D_+(\t a)D_-(\t a')\Big) \, .
\end{eqnarray}
In the limit of $\bet=0$, $\Om_\m=1$ and $\Om_\phi=0$ the kernel $\F_2$ reduces to the standard form of $\F_2$ in the E-dS Universe \cite{Bernardeau:2001qr}. Here we should mention that in the subhorizon approximation the first three curly bracketed terms of Eq.~(\ref{eq:K}) dominate over the other terms.


\section{Power Spectrum and Bispectrum}
\label{sec:power}

Power spectrum is the Fourier transform of the two point correlation function. It is one the important statistical quantities to describe the matter perturbation. The matter power spectrum $\P(\tau,k)$ is defined as
\begin{equation}
\Big<\del(\tau,\vck)\del(\tau,\vck')\Big>=\del^{(3)}(\vck+\vck')\P(\tau,k) \, ,
\end{equation}
where $\big<\dots\big>$ represents the ensemble average. Dependence of $\P(\tau,k)$ only on the values of $\vck$ and not on the vector $\vck$ is a consequence of the assumption of statistical homogeneity and isotropy of the initial fluctuations. 

Considering the growing mode solution of Eq.~(\ref{eq:den_con_evo}) the power spectrum can be written as \cite{Eisenstein:1997jh,Duniya:2015nva}
\begin{eqnarray}
 \P(\tau,k)&\propto& A_{\rm H}^2\Big|D_+(\t a)\Big|^2 T^2(k)\(\frac{k}{H_0}\)^{n_\s} \, ,
 \label{eq:PS}
\end{eqnarray}
where $A_{\rm H}$ is a normalization factor which can be fixed folowing the procedure discussed below and $n_\s$ is the spectral index of scalar perturbation during inflation. $D_+(\t a)$ is given in the Eq.~(\ref{eq:Dp})  and $T(k)$ is the transfer function \cite{Eisenstein:1997ik} which relates the primordial curvature perturbation with the comoving matter perturbation. We use the Eisenstein-Hu fitting formula for the transfer function \cite{Eisenstein:1997ik}.

Power spectrum defined in Eq.~(\ref{eq:PS}) has a dimension. One can also define the dimensionless power spectrum
\begin{eqnarray}
 \Delta^2(\tau,k)=\frac{k^3}{2\pi^2}\P(\tau,k) \, .
\end{eqnarray}

The rms amplitude of mass fluctuations $\sig$ is given by
\begin{eqnarray}
 \sig_{\rm R}^2=\int_0^\infty \d k \frac{k^2}{2\pi^2}\P(\tau,k)\big|W_{\rm win}(kR)\big|^2 \, ,
\end{eqnarray}
where $W_{\rm win}(kR)$ is the window function of size $R$ with which we define a smoothed density field
\begin{equation}
 \del(\vec{x};R)=\int\del(\vec{x}') W_{\rm win}(\vec{x}-\vec{x}';R) \d^3x' \, .
\end{equation}
Since the above relation is a convolution the Fourier transform of the smoothed density field is a product of $\del(\vck)$ and $W_{\rm win}(kR)$. We choose spherical top-hat window function which is given by
\begin{eqnarray}
 W_{\rm win}(kR)=\frac{3}{(kR)^3}\Big(\sin(kR)-kR \cos(kR)\Big) \, ,
\end{eqnarray}
The smoothing scale at which $\sig_{\rm R}\sim 1$ represents the scale at which the linear perturbation theory breaks and nonlinear effects become important. In this regard $R=8~h^{-1}\rm Mpc$ is a relevant scale and from Planck 2015 results we have, at present ($z=0$), $\sig_8=0.8159\pm 0086$ \cite{Ade:2015xua}. Using this best fit value of $\sig_8$ we fix the normalization factor $A_{\rm H}$ of the power spectrum ~(\ref{eq:PS}). The evolution of $\sig_8(z)$ can be represented by the growth function $D_+(z)$ as follows 
\begin{equation}
 \sig_8(z)=\sig_8(0)\frac{D_+(z)}{D_+(0)} \, .
\end{equation}
To fix the normalization we fix the value of $\sig_8(z)$, calculated in the scenario under consideration, same as in the $\Lam$CDM model at high redshift using the above equation and the fact that $\sig_8(z=0)=0.8159$ for $\Lam$CDM case.

\begin{figure}[h]
\centering
\includegraphics[scale=.9]{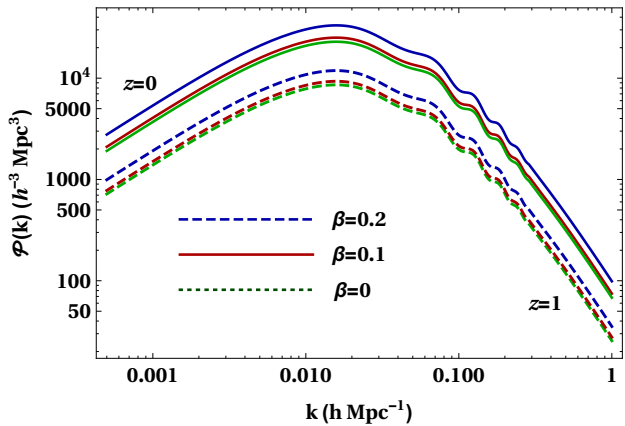}
\caption{Matter Power spectrum at redshift $z=0$ (solid lines) and $z=1$ (dashed lines) are plotted for the potential~\eqref{eq:pot1} with $\mu_1=20$, $\mu_2=0.1$. Green (lower solid and dashed), red (middle solid and dashed) and blue (top solid and dashed) curves are for $\bet=0,\; 0.1$ and $0.2$ respectively. We have taken $\Om_{\rm b}$, fractional energy density of baryon=0.04, $\Om_{\m0}=0.3$ and $n_\s=0.968$.} 
\label{fig:ps_dexp}
\end{figure}

In Fig.~\ref{fig:ps_dexp} the nature of the power spectrum has been shown at redshifts 0 and 1 for different values of $\bet$. Since we have normalized the $\sig_8$ at high redshift the nature of the power spectrum at $z=0$ changes for different $\bet$ unlike the case where the normalization is fixed at $z=0$. In Fig.~\ref{fig:ps_dexp} we can see that the power spectrum gets enhanced as the values of $\bet$ increase.

\begin{figure}[h]
\centering
\includegraphics[scale=.9]{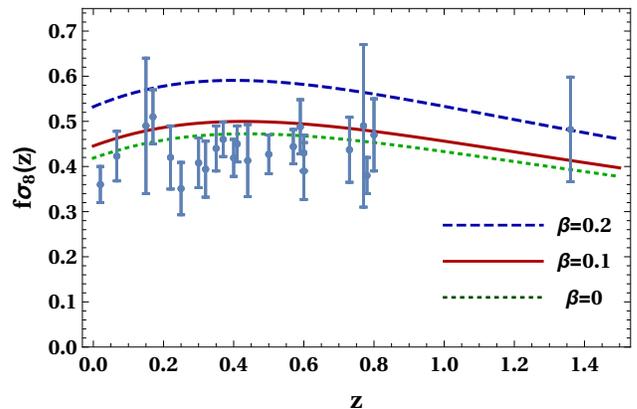}
\caption{$f\sig_8(z)$ is plotted for the potential~\eqref{eq:pot1} with $\mu_1=20$, $\mu_2=0.1$. Green (dotted), red (solid) and blue (dashed) lines are for $\bet=0,\; 0.1$ and $0.2$ respectively. Dots (blue) are the measured values of $f\sig_8(z)$ with 1$\sig$ error bars. Observationally measured values of $f\sig_8(z)$ are listed in the Table~\ref{tab:fsig8} in the Appendix~\ref{app_fsig8}.} 
\label{fig:fsig8_dexp}
\end{figure}

We can also define growth factor $f$ as
\begin{equation}
 f=\frac{\d \ln D_+}{\d \ln a} \, .
\end{equation}
The product of the growth factor and $\sig_8(z)$ {\it i.e.}, $f\sig_8(z)$ is a observationally measurable quantity. Fig.~\ref{fig:fsig8_dexp} compares the numerical evolution of $f\sig_8(z)$ with observational data of $f\sig_8(z)$ for different values of $\bet$ and it seems that the larger values of $\bet$ can be excluded from these data.

Another important statistical quantity is matter bispectrum which is related to the three point function through a Fourier transformation. This is also important for the mildly nonlinear evolution of the density fluctuations and non-Gaussianity. Matter bispectrum is given by the relation
\begin{eqnarray}
 \big<\del(\tau,\vck)\del(\tau,\vck')\del(\tau,\vck'')\big>=\del^{(3)}(\vck+\vck'+\vck'')\B(\tau,k,k'),~~~~~
\end{eqnarray}
where
\begin{eqnarray}
 \B(\tau,k,k')=2\F_2(\vck,\vck')\P(k)\P(k')+\rm cyc \, .
\end{eqnarray}

\begin{figure}[h]
\centering
\includegraphics[scale=.8]{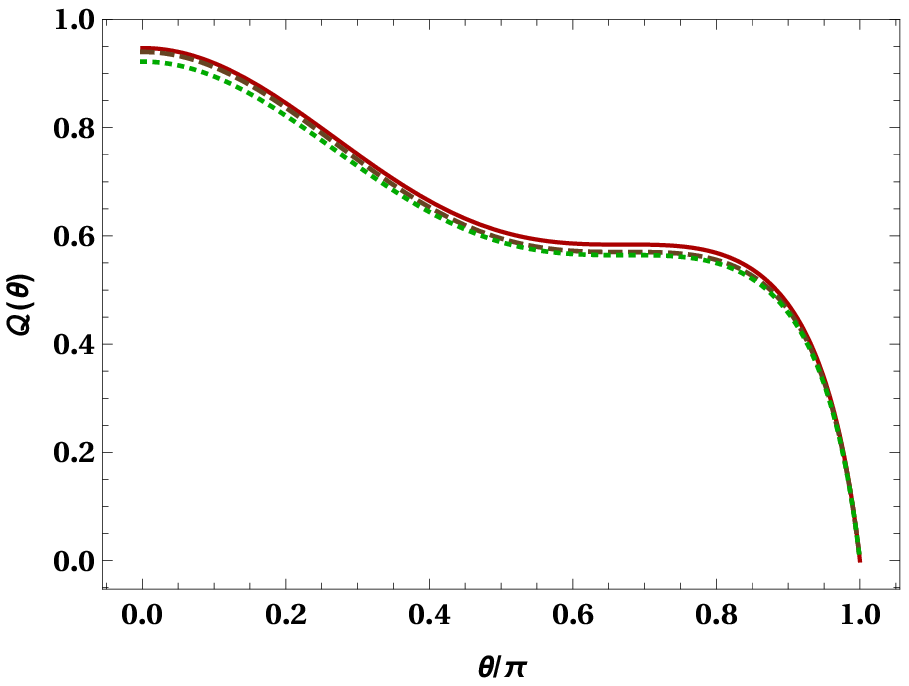}\vskip10pt
\includegraphics[scale=.8]{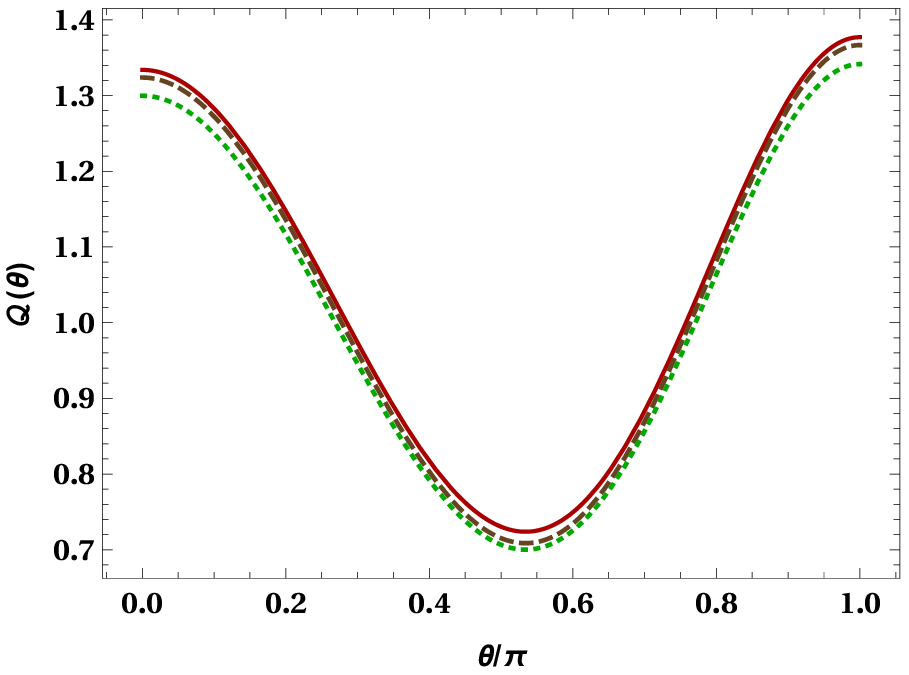}
\caption{Considering the potential~\eqref{eq:pot1} with $\mu_1=20$ and $\mu_2=0.1$ reduced bispectrum, as a function of the angle $\theta$, at $z=0$, is shown for $k=k'=0.01~\rm hMpc^{-1}$ (upper figure) and $5k=k'=0.05~\rm hMpc^{-1}$ (lower figure). In both the figure the brown (dashed), green (dotted) and red (solid) lines represent the bispectrum for $\bet=0,\; 0.5$ and $\Lam$CDM respectively.} 
\label{fig:bisp_dexp}
\end{figure}

It is more convenient to use the reduced bispectrum $\Q$ defined as
\begin{eqnarray}
\Q= \frac{\B(\tau,k,k')}{\P(\tau,k)\P(\tau,k')+\P(\tau,k')\P(\tau,k'')+\dots} \, ,
\end{eqnarray}
because it is scale and time independent to lowest order in nonlinear perturbation theory. Fig.~\ref{fig:bisp_dexp} shows the the nature of reduced bispectrum at $z=0$ and compares with that of the $\Lam$CDM case. We can see that there is not much difference between different curves of the reduced bispectrum. So the effect of the conformal coupling $\bet$ on the reduced bispectrum is insignificant.

\section{Summary and Conclusion}
\label{sec:conc}

In this work we studied the LMG scenario \cite{Ali:2012cv,Hossain:2012qm} at both background and perturbation level. LMG has cubic Galileon action with potential. Potential is added phenomenologically and responsible for the late time acceleration in this scenario. In the current study we have considered potentials which can lead to tracker behavior of the scalar field. In this regard we considered potentials~\eqref{eq:pot1} and \eqref{eq:pot2}. These potentials can lead to similar dynamics and we considered the potential~\eqref{eq:pot1} for the numerical purpose. 

At the perturbation level we have studied up to second order perturbation. The linear perturbation of LMG in subhorizon approximation was studied in Ref.~\cite{Ali:2012cv}. In this work we have calculated the integral solution of the growing and decaying modes in subhorizon approximation by generalizing some results of Ref.~\cite{Bartolo:2013ws} for conformal coupling and any potential. Here we have considered exponential form of the conformal factor with conformal coupling constant $\bet$ and have studied the effect of $\bet$ at the perturbation level. It is found that a simple transformation~\eqref{eq:atilde} reduces the  modified evolution equation of the density contrast~\eqref{eq:den_con_evo} to the standard form~\eqref{eq:den_con_evo1} with $\H$ replaced by $\t\H$ defined in Eq.~\eqref{eq:Htilde} and the Newton's constant, $G$, replaced by $\t G_{\rm eff}$ defined in Eq.~\eqref{eq:Gefftilde}. Eq.~\eqref{eq:den_con_evo1} is very useful to calculate the integral solutions of growing and decaying modes {\it i.e.}, Eqs.~\eqref{eq:Dp} and \eqref{eq:Dm} (for details of this calculation see Appendix~\ref{app_int}). The evolution equation for the second order density contrast (Eq.~\eqref{eq:den_con_evo2}) is same as the first order except one source term $\S_\del$ (Eq.~\eqref{eq:S_del}). The solution of Eq.~\eqref{eq:den_con_evo2} can be written in terms of the first order growing and decaying modes (Eq.~\eqref{eq:den2}). The kernel of the second order perturbation, $\F_2$, is also written with the help of growing and decaying modes which reduces to the E-dS kernel at some limit. 

Next we investigated matter power spectrum and bispectrum. The power spectrum is normalized in the past by considering the best fit value of $\sig_8(z)$ at present from  the current observation \cite{Ade:2015xua}, {\it i.e.}, $\sig_8(z=0)=0.8159$ for $\Lam$CDM. The power spectrum changes for different $\bet$. As $\bet$ increases power spectrum gets enhanced (see Fig.~\ref{fig:ps_dexp}). So the study of the power spectrum can tell us about the allowed values of $\bet$. We used RSD data to compare the effects of different values of $\bet$ by calculating $f\sig_8(z)$. The comparison is depicted in Fig.~\ref{fig:fsig8_dexp}, from which we can see that as we increase the values of $\bet$ it is more probable for the $\bet$ to be excluded by the RSD data. Finally we studied the matter bispectrum. The reduced bispectrum, for two different combinations of $k$ and $k'$, is plotted in Fig.~\ref{fig:bisp_dexp} and the dependence on $\bet$ is very small. In this study we did not do the full statistical analysis to obtain the bound on $\bet$, which we expect to do in future.

In summary, in this paper, we studied the effect of the conformal coupling at the perturbation level in a tracker scalar field model with a cubic Galileon correction term. Here we should mention that the effects we can observe in Figs.~\ref{fig:ps_dexp} and \ref{fig:fsig8_dexp} are from both the conformal coupling and the cubic Galileon term. This means that if we put $\al=0$, {\it i.e.}, without cubic Galileon term, the standard tracker scalar field may not exactly reproduce the same effect with different values of $\bet$. In other words, for the same $\bet$ the scenario under consideration has different effects than the standard tracker scalar field. But these effects are not significant and within the $1\sig$ error bars of $f\sig_8$ data. In future, if we get more precise data then it may be possible to distinguish between the LMG and the standard tracker scalar field.


\begin{acknowledgments}
The author thanks Emilio Bellini, Antonio De Felice, Bikash R. Dinda, Jinn-Ouk Gong, Aseem Paranjape and Mohammad Sami for useful discussions. The comments on the manuscript by Jinn-Ouk Gong and Mohammad Sami are gratefully acknowledged.
\end{acknowledgments}


\appendix

\section{First order perturbation and subhorizon approximation}
\label{app_sub}

First order perturbation of Einstein tensor gives,
\begin{align}
 \delta G^{~~~0}_{(1)0} =&\frac{2}{a^2}\(3\H(\H\Phi_1+\dot\Psi_1)-\nabla^2\Psi_1\) \, ,\\
 \delta G^{~~~0}_{(1)i} =& -\frac{2}{a^2}\partial_i\(\dot\Psi_1+\H\Phi_1\)=-\delta G^{~~~i}_{(1)0} \, , \\
 \delta G^{~~~i}_{(1)j} =& \frac{1}{a^2}\Bigg(2\ddot\Psi_1+2\H\dot\Phi_1+4\H\dot\Psi_1+2(2\dot\H+\H^2)\Phi_1 \nn \\* & +\nabla^2(\Phi_1-\Psi_1)\Bigg)\delta^i_j-
\frac{1}{a^2}\partial_i\partial_j(\Phi_1-\Psi_1) \, .
\end{align}

In subhorizon approximation and for the linear perturbation
\begin{eqnarray}
 \frac{2k^2\Psi_1}{a^2} && \approx -\frac{1}{\Mpl^2}\bigg(\del_1\bar\rho_\m+\del\rho_\phi^{(1)}\bigg) \, , \\ 
\del\rho_\phi^{(1)} &&\approx -\Mpl Q\frac{k^2\del\phi_1}{a^2} \, ,
\label{approx} \\
\delta p_\phi^{(1)} &&\approx 0\, ,\\
\frac{k^2\Phi_1}{a^2} && \approx -\frac{P+\bet Q}{\Mpl^2(2P-Q^2)}\del_1\bar\rho_\m = \frac{k^2\Psi_1}{a^2}\, ,\\
\frac{k^2\del\phi_1}{a^2} && \approx -\frac{Q+2\bet}{\Mpl(2P-Q^2)}\del_1\bar\rho_\m \, .
\end{eqnarray}


\section{Integral solutions}
\label{app_int}

Let us define a function $u(\t a)$ such as
\begin{equation}
 \del_1(\t a)=\frac{u(\t a)}{\t a} \sqrt{\frac{\t\H_0}{\t\H}} \, .
\end{equation}

In terms of $u(\t a)$ Eq.~(\ref{eq:den_con_evo12}) can be written as 
\begin{equation}
 \frac{\d^2u(\t a)}{\d \t a^2}-I(\t a)u(\t a)=0 \, ,
 \label{eq:ua}
\end{equation}
where
\begin{equation}
 I(\t a)= A(\t a)+\frac{1}{\t a\t\H}\frac{\d \t\H}{\d \t a}-\frac{1}{4\t\H^2}\(\frac{\d \t\H}{\d\t a}\)^2+\frac{1}{2\t\H}\frac{\d^2\t\H}{\d\t a^2} \, .
\end{equation}

Now consider the following differential equation
\begin{equation}
 \frac{\d^2y(\t a)}{\d \t a^2}+\frac{\d}{\d \t a}\Big(y(\t a)g(\t a)\Big)=0 \, .
 \label{eq:ya1}
\end{equation}
Using the transformation
\begin{equation}
 Y(\t a)=y(\t a)\e^{\frac{1}{2}\int_{\t a_\m}^{\t a} \d\t a' g(\t a')} \, ,
\end{equation}
with $\t a_\m$ as an initial value of $\t a$, we can have
\begin{equation}
 \frac{\d^2Y(\t a)}{\d\t a^2}+\frac{1}{2}\(\frac{\d g}{\d\t a}-\frac{1}{2}g^2(\t a)\)Y(\t a)=0 \, .
 \label{eq:Ya}
\end{equation}
Now, $g(\t a)$ can be a solution of
\begin{equation}
 \frac{\d g(\t a)}{\d\t a}-\frac{1}{2}g^2(\t a)+2 I(\t a)=0\, ,
 \label{eq:gad}
\end{equation}
which makes the Eq.~(\ref{eq:ua}) and (\ref{eq:Ya}) same and we can write
\begin{equation}
 u(\t a)=y(\t a)\e^{\frac{1}{2}\int_{\t a_\m}^{\t a} \d\t a' g(\t a')} \, ,
\end{equation}
and 
\begin{equation}
 \del_1(\t a)=\frac{y(\t a)}{\t a}\sqrt{\frac{\t\H_0}{\t\H}} ~ \e^{\frac{1}{2}\int_{\t a_\m}^{\t a} \d\t a' g(\t a')} \, .
 \label{eq:del_3}
\end{equation}

Now solving Eq.~(\ref{eq:ya1}) we obtain \cite{Bartolo:2013ws}
\begin{equation}
 y(\t a)=\gam^2(\t a)\(\kappa_1+\kappa_2\int_{\t a_\m}^{\t a}\frac{\d \t a'}{\gam^2(\t a')}\) \, ,
\end{equation}
where
\begin{equation}
 \gam^2(\t a)=\e^{-\int_{\t a_\m}^{\t a} \d\t a' g(\t a')} \, .
 \label{eq:gam_ai}
\end{equation}

Now putting the value of $y(\t a)$ in Eq.~(\ref{eq:del_3}) we get two independent solutions
\begin{eqnarray}
 D_1(\t a)&=&\frac{\gam(\t a)}{\t a}\sqrt{\frac{\t\H_0}{\t\H}}\, , 
 \label{eq:D1}\\
 D_2(\t a)&=&\frac{\gam(\t a)}{\t a}\sqrt{\frac{\t\H_0}{\t\H}}\int_{\t a_\m}^{\t a}\frac{\d \t a'}{\gam^2(\t a')} \, , 
 \label{eq:D2}
\end{eqnarray}
of
\begin{eqnarray}
  \delta_1(\t a)&=& \kappa_1 D_1(\t a)+\kappa_2 D_2(\t a) \, .
\end{eqnarray}

To Calculate the growing and decaying modes we shall consider the conventional normalization where we take $D_+(a)\sim a$ and $D_-(a)\sim \H/a\sim a^{-3/2}$ during the matter dominated era where the effect of the dark energy is negligible. During matter domination if we impose $D_1(a)=D_+(a)=a$, then we obtain 
\begin{eqnarray}
 g(\t a)&=&\frac{1}{\t a}\bigg[-\frac{7}{2}+\frac{\bet}{\Mpl}\frac{1}{\A(\phi)}\(\frac{3}{2}\frac{\d \phi}{\d \ln a} \nn \right.\\* && \left.-\frac{1}{\A(\phi)}\frac{\d^2 \phi}{\d \ln a^2} \)\bigg]\, . 
 \label{eq:ga} \\
 \gam(\t a)&=&\(\frac{\t a}{\t a_\m}\)^{7/4}\e^{-\frac{3\bet(\phi-\phi_\m)}{4\Mpl}}\(\frac{\A(\phi)}{\A(\phi_\m)}\)^{1/2} ,
 \label{eq:gam_a} \\
 \A(\phi)&=&\frac{\d \ln\t a}{\d \ln a}=1+\frac{\bet}{\Mpl}\frac{\d \phi}{\d \ln a} \, ,
 \label{eq:Aphi}
\end{eqnarray}
where $\t a_\m$ is the value of $\t a$ at some initial time $\tau_\m$ (or $a_\m$) during the matter domination and $\phi_\m=\phi(\tau_\m)$. Eq.~(\ref{eq:ga}) sets the initial condition for the Eq.~(\ref{eq:gad}) during the matter dominated era. The minimal case {\it i.e.}, $\bet=0$ gives $\t a=a$ and $g(a)=-7/(2a)$ and $\gam(a)=(a/a_\m)^{7/4}$\cite{Bartolo:2013ws}. 

Now to normalize the decaying mode we need to perform the integration $\int_{\t a_\m}^{\t a}\d \t a'/\gam^2(\t a')$ during the matter dominated era. To do this we shall consider an approximation. Since during matter domination the scalar field is subdominant we consider the value of the term $\e^{\bet\phi/\Mpl}$ close to $\e^{\bet\phi_\m/\Mpl}$. Considering this approximation we obtain
\begin{eqnarray}
 \int_{\t a_\m}^{\t a}\frac{\d \t a'}{\gam^2(\t a')} &=& \t a_\m^{7/2}\e^{-3\bet\phi_\m/(2\Mpl)}\A(\phi_\m) \int_{a_\m}^{a}\frac{\d a'}{a'^{7/2}\e^{\bet\phi/\Mpl}} \nn \\* &=& -\frac{2}{5}\t a_\m^{7/2}\e^{-5\bet\phi_\m/(2\Mpl)}\A(\phi_\m) \nn \\* && \times \(a^{-5/2}-a_\m^{-5/2}\)~. ~~~~~~~~
\end{eqnarray}
Final forms, with normalization, of the growing and decaying modes are given in Eqs.~\eqref{eq:Dp} and \eqref{eq:Dm} respectively. 


\section{Second order perturbation and subhorizon approximation}
\label{app_2nd}

In second order perturbation in addition to the linear terms of second order perturbation there will be nonlinear terms consists of products of two first order perturbation terms. Second order perturbation of Einstein tensor gives
\begin{align}
 \delta G^{~~~0}_{(2)0} =&\frac{2}{a^2}\(3\H(\H\Phi_2+\dot\Psi_2)-\nabla^2\Psi_2\) - S_{\rm G00}\, ,\\
 \delta G^{~~~0}_{(2)i} =& -\frac{2}{a^2}\partial_i\(\dot\Psi_2+\H\Phi_2\)-S_{\rm G0i} \, , \\
 \delta G^{~~~i}_{(2)j} =& \frac{1}{a^2}\Bigg(2\ddot\Psi_2+2\H\dot\Phi_2+4\H\dot\Psi_2+2(2\dot\H+\H^2)\Phi_2 \nn \\* & +\nabla^2(\Phi_2-\Psi_2)\Bigg)\delta^i_j-
\frac{1}{a^2}\partial^i\partial_j(\Phi_2-\Psi_2) \nn \\* & -S_{\rm G ij} \, ,
\end{align}
where,
\begin{eqnarray}
 S_{\rm G00} &=& \frac{2}{a^2}\bigg(3\dot\Psi_1^2+12\H(\Phi_1-\Psi_1)\dot\Psi_1+3\partial_i\Psi_1\partial^i\Psi_1 \nn \\* && +12\H^2\Phi_1^2+8\Psi_1\nabla^2\Psi_1\bigg) \, , 
 \label{eq:sg00}
 \end{eqnarray}
 \begin{eqnarray}
 S_{\rm G0i} &=& -\frac{4}{a^2}\bigg\{\partial_i\Big((\Phi_1-2\Psi_1)\dot\Psi_1\Big)+\Phi_1\partial_i\Big(\dot\Psi_1 \nn \\ &&  +4\H\Phi_1\Big)\bigg\} \, , 
 \label{eq:sgi0}
\end{eqnarray}
\begin{eqnarray}  
 S_{\rm Gij} &=& \frac{2}{a^2}\Bigg[\Bigg\{2(\Phi_1-\Psi_1)\nabla^2\Phi_1+4\Psi_1\nabla^2\Psi_1+4\big(2\dot\H \nn \\ &&+\H^2\big)\Phi_1^2+8\H(\Phi_1-\Psi_1)\dot\Psi_1+2(\dot\Psi_1+4\H\Phi_1)\dot\Phi_1 \nn \\ && +4(\Phi_1-\Psi_1)\ddot\Psi_1  -\dot\Psi_1^2+\partial_k\Phi_1\partial^k\Phi_1 \nn \\ &&+2\partial_k\Psi_1\partial^k\Psi_1\Bigg\}\del_j^i-2(\Phi_1-\Psi_1)\partial^i\partial_j\Phi_1  \nn \\ && +\partial^i\Phi_1\partial_j\Psi_1+\partial^i\Psi_1\partial_j\Phi_1-\partial^i\Phi_1\partial_j\Phi_1 \nn \\ && -4\Psi_1\partial^i\partial_j\Psi_1  -3\partial^i\Psi_1\partial_j\Psi_1\Bigg] \, .
 \label{eq:sgij}
\end{eqnarray}

Second order terms of a perturbed general energy momentum tensor ($T_{\mu\nu}$) are
\begin{eqnarray}
 \del T^{~~~0}_{(2)0} &=& -\del\rho_2-2(\bar\rho+\bar p)\partial_i\ups_1\partial^i\ups_1 \, , \\
 \del T^{~~~0}_{(2)i} &=& -(\bar\rho+\bar p)\partial_i\ups_2-2(\del\rho_1+\del p_1)\partial_1\ups_1 \nn \\ && +2(\bar\rho+\bar p)(\Phi_1+2\Psi_1)\partial_i\ups_1 \, ,\\
 \del T^{~~~i}_{(2)j} &=& \del p_2\del^i_j+2(\bar\rho+\bar p)\partial^i\ups_1\partial_j\ups_1 \, ,
\end{eqnarray}
where $\bar\rho$ and $\bar p$ are the background energy density and pressure.

Second order perturbation of the energy momentum tensor of LMG gives
\begin{widetext}
 \begin{eqnarray}
  \delta T^{(\phi)0}_{(2)~0} &=& -\frac{1}{a^2}\bigg[\dot{\phi}\(1-\frac{9\al}{M^3a^2}\H\dot{\phi}\)\dot{\del\phi_2}+
a^2V'(\phi)\del\phi_2 +\frac{\dot{\phi}^2\al}{a^2M^3}\nabla^2\del\phi_2-\dot{\phi}^2\(1-\frac{12\al}{M^3a^2}\H\dot\phi\)\Phi_2
+\frac{3\al}{M^3 a^2}\dot{\phi}^3\dot\Psi_2 \bigg] \nn \\* && +S_{\T00} \, , \\ \nn \\
S_{\T00} &=& -\frac{1}{a^2}\bigg[\(\dot{\del\phi_1}^2-4\dot\phi\(\dot{\del\phi_1}-\Phi_1\dot\phi\)\Phi_1\)\(1-\frac{18\al}{M^3a^2}\H\dot\phi\)+\partial_i\del\phi_1\partial^i\del\phi_1\(1+\frac{2\al}{M^3a^2}\H\dot\phi\) +a^2V''(\phi)\del\phi_1^2  \nn \\ && +\frac{\al}{M^3a^2}\bigg\{4\dot\phi\(\dot{\del\phi_1}-\dot\phi(\Phi_1-\Psi_1)\)\nabla^2\del\phi_1+6\dot\phi^2\dot\Psi_1\(3\dot{\del\phi_1}-2\dot\phi(2\Phi_1-\Psi_1)\)-2\dot\phi^2\partial_i\del\phi_1\partial^i\Psi_1\bigg\}\bigg] \, ;
\label{eq:st00}
 \end{eqnarray}
 
 \begin{eqnarray}
  \delta T^{(\phi)0}_{(2)~i} &=& -\frac{\dot{\phi}}{a^2}\partial_i\bigg[\frac{\al}{M^3a^2}\dot{\phi}\dot{\del\phi_2}+\del\phi_2\left(1-\frac{3\al}{M^3a^2}\H\dot{\phi} \right)  -\frac{\al}{M^3a^2}\dot{\phi}^2\Phi_2\bigg] +S_{\T 0i} \, , \\ \nn \\
  S_{\T 0i}&=& \frac{2}{a^2}\Bigg[\(2\dot\phi\Phi_1-\dot{\del\phi_1}\)\partial_i\delta\phi_1+\frac{\al}{M^3a^2}\Bigg\{\dot\phi\(2\dot\phi\Phi_1-\dot{\del\phi_1}\)\(2\partial_i\dot{\del\phi_1}-3\dot\phi\partial_i\Phi_1\)+\dot\phi\Bigg(6\H\(\dot{\del\phi_1}-\dot\phi\Phi_1\)-3\dot\phi\dot\Psi_1 \nn \\* && -\nabla^2\del\phi_1+\partial_i\partial^i\del\phi_1\Bigg)\partial_i\del\Phi_1+\dot\phi\(\partial_i\partial^j\del\phi_1\partial_j\del\phi_1+\partial_i\partial^k\del\phi_1\partial_k\del\phi_1\)\Bigg\}\Bigg] \, ;
 \label{eq:tpi02}
 \end{eqnarray}

 \begin{eqnarray}
  \delta T^{(\phi)i}_{(2)~j} &=&\frac{1}{a^2}\bigg[\frac{\al}{M^3a^2}\dot{\phi}^2\ddot{\del\phi_2}+\dot{\phi}\dot{\del\phi_2}\(1+\frac{\al}{M^3a^2}\(2\ddot{\phi}-3\H\dot\phi\)\)-a^2 V'(\phi) \del\phi_2 -\dot{\phi}^2\Phi_2 \(1+\frac{4\al}{M^3a^2}\(\ddot{\phi}-\H\dot\phi\)\) \nn \\ && -\frac{\al}{M^3a^2}\dot{\phi}^3\dot{\Phi}_2\bigg]\delta^i_j +S_{\T ij} \, ,\\ \nn \\
  S_{\T ij} &=& \frac{1}{a^2}\bigg[\frac{4\al}{M^3a^2}\dot\phi\(\dot{\del\phi_1}-2\dot\phi\Phi_1\)\ddot{\del\phi_1}+\bigg\{\(1+\frac{2\al}{M^3a^2}\(\ddot\phi-3\H\dot\phi\)\)\dot{\del\phi_1}-\frac{6\al}{M^3a^2}\dot\phi^2\dot\Phi_1 \nn \\ &&-4\dot\phi\(1-\frac{2\al}{M^3a^2}\(3\H\dot\phi-2\ddot\phi\)\)\Phi_1\bigg\}\dot{\del\phi_1} -\frac{1}{3}\(1+\frac{2\al}{M^3a^2}\(\ddot\phi-\H\dot\phi\)\)\partial_i\del\phi_1\partial^i\del\phi_1-\frac{8\al}{3M^3a^2}\dot\phi\partial_i\del\phi_1\partial^i\del\dot\phi_1 \nn \\ &&+\frac{2\al}{3M^3a^2}\dot\phi^2\partial_i\del\phi_1\partial^i\Phi_1+4\dot\phi^2\(1+\frac{6\al}{M^3a^2}\(\ddot\phi-\H\dot\phi\)\)\Phi_1^2+\frac{12\al}{M^3a^2}\dot\phi^3\Phi_1\dot\Phi_1-a^2V''(\phi)\del\phi_1^2\bigg]\del_j^i \, .
  \label{eq:stij}
 \end{eqnarray}
 
 Second order perturbation of the LMG field's equation of motion reads
 \begin{eqnarray}
&& \left(1-\frac{6\alpha}{M^3a^2}\H\dot{\phi}\right)\ddot{\del\phi_2}+2\left(\H- \frac{3\al}{M^3a^2}\(\H\ddot{\phi}+\dot{\phi}\dot{\H}\)\right)\dot{\delta\phi_2}+ a^2 V''(\phi)\del\phi_2 - \left(1-\frac{2\al}{M^3a^2}(\ddot{\phi}+\H\dot{\phi}) \right) \nabla^2\del\phi_2 \nn \\ && - \dot\phi\left(1-\frac{9\al}{M^3a^2}\H\dot\phi\right)\dot\Phi_2 +
2\bigg[\(V'(\phi)+\frac{\bet}{\Mpl}\rho_\m\)a^2- \frac{3\al\dot\phi}{M^3a^2}\left(2\H\ddot{\phi}+\dot{\phi}\dot\H\right)\bigg] \Phi_2+ \frac{\al}{M^3a^2}\dot\phi^2\nabla^2\Phi_2 + \frac{3\al}{M^3a^2}\dot\phi^2\ddot\Psi_2 \nn \\  && -3\dot\phi\left(1-\frac{3\al}{M^3a^2}(2\ddot{\phi}+ 3\H\dot{\phi})\right)\dot\Psi_2  =-\(\frac{\bet(\phi)}{\Mpl}\del_2+\frac{\bet'(\phi)}{\Mpl}\del\phi_2\)a^2\bar\rho_\m+S_{\rm eom} \, ,
\label{pieom2}
\end{eqnarray}

\begin{eqnarray}
S_{\rm eom} &=& 4\(1-\frac{12\al}{M^3a^2}\H\dot\phi\)\Phi_1\ddot{\del\phi_1}+2\(1-\frac{18\al}{M^3a^2}\H\dot\phi\)\dot\Phi_1\dot{\del\phi_1}+8\(\H-\frac{6\al}{M^3a^2}(\dot\H\dot\phi+\H\ddot\phi)\)\Phi_1\dot{\del\phi_1} \nn \\ && +6\(1-\frac{2\al}{M^3a^2}(3\H\dot\phi+\ddot\phi)\)\dot\Psi_1\dot{\del\phi_1} +4\(1-\frac{2\al}{M^3a^2}(\H\dot\phi+\ddot\phi)\)\Psi_1\nabla^2\del\phi_1 -4\dot\phi\(2-\frac{27\al}{M^3a^2}\H\dot\phi\)\Phi_1\dot\Phi_1 \nn \\ && -12\dot\phi\(1-\frac{2\al}{M^3a^2}(3\H\dot\phi+2\ddot\phi)\)\Phi_1\dot\Psi_1+8\(\Big(V'(\phi)+\frac{\bet}{\Mpl}\bar\rho_\m\Big)a^2+\frac{6\al}{M^3a^2}\dot\phi(\dot\H\dot\phi+2\H\ddot\phi)\)\Phi_1^2 \nn \\ && +12\dot\phi\(1-\frac{\al}{M^3a^2}(3\H\dot\phi+2\ddot\phi)\)\Psi_1\dot\Psi_1+2\partial_i\del\phi_1\partial^i\Phi_1-2\partial_i\del\phi_1\partial^i\Psi_1+\frac{\al}{M^3a^2}\Bigg\{\dot\phi^2\Bigg(12(2\Phi_1-\Psi_1)\ddot\Psi_1+6\dot\Psi_1^2 \nn \\ && +18\dot\Phi_1\dot\Psi_1+4(2\Phi_1-\Psi_1)\nabla^2\Phi_1\Bigg)-8(\ddot\phi+\H\dot\phi)\Phi_1\nabla^2\del\phi_1-4(\ddot{\del\phi_1}+\H\dot{\del\phi_1})\nabla^2\del\phi_1+12\H\dot{\del\phi_1}\ddot{\del\phi_1}+6\dot\H\dot{\del\phi_1}^2 \nn \\ && -4\dot\phi\dot{\del\phi_1}\nabla^2\Phi_1  +4\dot\phi(\dot\Phi_1+2\dot\Psi_1)\nabla^2\del\phi_1-12\dot\phi\dot\Psi_1\ddot{\del\phi_1}-12\dot\phi\dot{\del\phi_1}\ddot\Psi_1+4\partial_i\dot{\del\phi_1}\partial^i\dot{\del\phi_1}-8\H\partial_i\dot{\del\phi_1}\partial^i\del\phi_1 \nn \\ && -8\dot\phi\partial_i\Phi_1\partial^i\dot{\del\phi_1}+2\(\nabla^2\del\phi_1\)^2-2\partial_i\partial_j\del\phi_1\partial^i\partial^j\del\phi_1+6\dot\phi^2\partial_i\Phi_1\partial^i\Phi_1+2\dot\phi^2\partial_i\Phi_1\partial^i\Psi_1-2\dot\H\partial_i\del\phi_1\partial^i\del\phi_1  \nn \\ && +8\dot\phi\partial_i\dot\Psi_1\partial^i\del\phi_1 +4\H\dot\phi\partial_i\Phi_1\partial^i\del\phi_1+4\H\dot\phi\partial_i\Psi_1\partial^i\del\phi_1+4\ddot\phi\partial_i\Psi_1\partial^i\del\phi_1\Bigg\}-2\frac{\bet'(\phi)}{\Mpl}a^2\bar\rho_\m \del_1\del\phi_1-\bigg(V'''(\phi) \nn \\ && +\frac{\bet''(\phi)}{\Mpl}\bar\rho_\m\bigg)a^2\del\phi_1^2 \, .
\label{eq:seom}
\end{eqnarray}
\end{widetext}

In subhorizon approximation
\begin{eqnarray}
 \frac{2k^2\Psi_2}{a^2} && \approx -\frac{1}{\Mpl^2}\bigg(\del_2\bar\rho_\m+\del\rho_\phi^{(2)}\nn \\* && +2(\bar\rho_\m+\bar\rho_\phi+\bar p_\phi)\partial_i\ups_1\partial^i\ups_1\bigg) \, , \\ 
\del\rho_\phi^{(2)} &&\approx -\Mpl Q\frac{k^2\del\phi_2}{a^2}-S_{\rm T00} \, ,
\label{approx} \\
\delta p_\phi^{(2)} &&\approx S_{\rm Tii}  \, ,\\
\frac{k^2\Phi_2}{a^2} && \approx -\frac{1}{\Mpl(2P-Q^2)}\bigg[\frac{P+\bet Q}{\Mpl}\del_2\bar\rho_\m-\frac{Q}{a^2}S_{\rm eom} ~~~\nn \\* && -\Mpl P S_{\del1}\bigg] \, ,\\
\frac{k^2\del\phi_2}{a^2} && \approx -\frac{1}{(2P-Q^2)}\bigg[\frac{Q+2\bet}{\Mpl}\del_2\bar\rho_\m-\frac{2}{a^2}S_{\rm eom} \nn \\ && -\Mpl Q S_{\del1}\bigg] \, .
\end{eqnarray}
where $S_{\del1}$ is defined in the Eq.~(\ref{eq:sdel1}).


\section{$f\sig_8$ data}
\label{app_fsig8}

Available data of $f\sig_8(z)$ is listed in the Table~\ref{tab:fsig8} \cite{DeFelice:2016ufg}.

\begin{table}
\caption{$f\sig_8$ data.}
\begin{center}
\label{tab:fsig8}
\begin{tabular}{ccc}
\hline\hline
~~$z$ & ~~~~$f\sig_8(z)$& ~~ Reference\\
\tableline
0.02  & 0.360  $\pm$  0.04 & \cite{Hudson:2012zga} \\
0.067  &  0.423  $\pm$  0.055 & \cite{Beutler:2012px} \\
0.15  & 0.490  $\pm$  0.15 & \cite{Howlett:2014opa} \\
0.17  &  0.510  $\pm$  0.06 & \cite{Percival:2004fs,Song:2008qt} \\
0.22  &  0.420  $\pm$  0.07 & \cite{Blake:2011rj} \\
0.25  &  0.351  $\pm$  0.058 & \cite{Samushia:2011cs} \\
0.3  &  0.408  $\pm$  0.0552 & \cite{Tojeiro:2012rp} \\
0.32  &  0.394  $\pm$  0.062 & \cite{Gil-Marin:2015sqa} \\
0.35  &  0.440  $\pm$  0.05 &  \cite{Song:2008qt,Tegmark:2006az} \\
0.37  & 0.460  $\pm$  0.038 & \cite{Samushia:2011cs} \\
0.4  &  0.419  $\pm$  0.041 &  \cite{Tojeiro:2012rp} \\
0.41  &  0.450  $\pm$  0.04 &  \cite{Blake:2011rj} \\
0.44  &  0.413  $\pm$  0.08 &  \cite{Blake:2012pj} \\
0.5  &  0.427  $\pm$  0.043 &  \cite{Tojeiro:2012rp} \\
0.57  &  0.444  $\pm$  0.038 & \cite{Gil-Marin:2015sqa} \\
0.59  &  0.488  $\pm$  0.06 & \cite{Chuang:2013wga} \\
0.6  &  0.430  $\pm$  0.04 & \cite{Blake:2011rj} \\
0.6 & 0.390    $\pm$  0.063 & \cite{Macaulay:2013swa} \\
0.73  &  0.437  $\pm$  0.072 & \cite{Macaulay:2013swa} \\
0.77  & 0.490  $\pm$  0.18 &  \cite{Guzzo:2008ac,Song:2008qt} \\
0.78  &  0.380  $\pm$  0.04 &  \cite{Blake:2011rj} \\
0.8  &  0.470 $\pm$  0.08 &  \cite{delaTorre:2013rpa} \\
1.36  &  0.482  $\pm$  0.116 & \cite{Okumura:2015lvp} \\
\hline\hline
\end{tabular}
\end{center}
\end{table}


\begin{thebibliography}{150}


\bibitem{Riess:1998cb} 
  A.~G.~Riess {\it et al.} [Supernova Search Team],
  \href{\doi/10.1086/300499}{Astron.\ J.\  {\bf 116}, 1009 (1998)}
  [\href{\arxiv/astro-ph/9805201}{astro-ph/9805201}].
  
\bibitem{Perlmutter:1998np} 
  S.~Perlmutter {\it et al.} [Supernova Cosmology Project Collaboration],
  \href{\doi/10.1086/307221}{Astrophys.\ J.\  {\bf 517}, 565 (1999)}
  [\href{\arxiv/astro-ph/9812133}{astro-ph/9812133}].
  
\bibitem{Ade:2015xua} 
  P.~A.~R.~Ade {\it et al.} [Planck Collaboration],
  \href{\doi/10.1051/0004-6361/201525830}{Astron.\ Astrophys.\  {\bf 594}, A13 (2016)}
  [arXiv:1502.01589 [astro-ph.CO]].



\bibitem{Copeland:2006wr} 
  E.~J.~Copeland, M.~Sami and S.~Tsujikawa,
  \href{\doi/10.1142/S021827180600942X}{Int.\ J.\ Mod.\ Phys.\ D {\bf 15}, 1753 (2006)}
  [\href{\arxiv/hep-th/0603057}{hep-th/0603057}].
  
\bibitem{Linder:2008pp} 
  E.~V.~Linder,
  \href{\doi/10.1088/0034-4885/71/5/056901}{Rept.\ Prog.\ Phys.\  {\bf 71}, 056901 (2008)}
  [\href{\arxiv/arXiv:0801.2968}{arXiv:0801.2968} [astro-ph]].
  
\bibitem{Silvestri:2009hh} 
  A.~Silvestri and M.~Trodden,
  \href{\doi/10.1088/0034-4885/72/9/096901}{Rept.\ Prog.\ Phys.\  {\bf 72}, 096901 (2009)}
  [\href{\arxiv/arXiv:0904.0024}{arXiv:0904.0024} [astro-ph.CO]].
  
\bibitem{Sahni:1999gb} 
  V.~Sahni and A.~A.~Starobinsky,
  \href{\doi/10.1142/S0218271800000542}{Int.\ J.\ Mod.\ Phys.\ D {\bf 9}, 373 (2000)}
  [\href{\arxiv/astro-ph/9904398}{astro-ph/9904398}].
  

  
\bibitem{Wetterich:1987fk} 
  C.~Wetterich,
  \href{\doi/10.1016/0550-3213(88)90192-7}{Nucl.\ Phys.\ B {\bf 302}, 645 (1988)}.
  
\bibitem{Wetterich:1987fm}
  C.~Wetterich,
  \href{\doi/10.1016/0550-3213(88)90193-9}{Nucl.\ Phys.\ B {\bf 302}, 668 (1988)}.

\bibitem{Ratra:1987rm}
  B.~Ratra and P.~J.~E.~Peebles,
  \href{\doi/10.1103/PhysRevD.37.3406}{Phys.\ Rev.\ D {\bf 37}, 3406 (1988)}.
  


\bibitem{Zlatev:1998tr} 
  I.~Zlatev, L.~M.~Wang and P.~J.~Steinhardt,
  \href{\doi/10.1103/PhysRevLett.82.896}{Phys.\ Rev.\ Lett.\  {\bf 82}, 896 (1999)}
  [\href{\arxiv/astro-ph/9807002}{astro-ph/9807002}].

\bibitem{Steinhardt:1999nw} 
  P.~J.~Steinhardt, L.~M.~Wang and I.~Zlatev,
  \href{\doi/10.1103/PhysRevD.59.123504}{Phys.\ Rev.\ D {\bf 59}, 123504 (1999)}
  [\href{\arxiv/astro-ph/9812313}{astro-ph/9812313}].
  
\bibitem{Scherrer:2007pu} 
  R.~J.~Scherrer and A.~A.~Sen,
  \href{\doi/10.1103/PhysRevD.77.083515}{Phys.\ Rev.\ D {\bf 77}, 083515 (2008)}
  [\href{\arxiv/arXiv:0712.3450}{arXiv:0712.3450} [astro-ph]].
  
\bibitem{Chiba:2009sj} 
  T.~Chiba,
  \href{\doi/10.1103/PhysRevD.80.109902}{Phys.\ Rev.\ D {\bf 79}, 083517 (2009)}
  Erratum: [\href{\doi/10.1103/PhysRevD.79.083517}{Phys.\ Rev.\ D {\bf 80}, 109902 (2009)}]
  [\href{\arxiv/arXiv:0902.4037}{arXiv:0902.4037} [astro-ph.CO]].
  


\bibitem{Fujii_Maeda} Y. Fujii and K. Maeda,
{\it The Scalar-Tensor Theory of Gravitation}
(Cambridge Monographs on Mathematical Physics, Cambridge Univ. Press, 2003).

\bibitem{Clifton:2011jh} 
  T.~Clifton, P.~G.~Ferreira, A.~Padilla and C.~Skordis,
  \href{\doi/10.1016/j.physrep.2012.01.001}{Phys.\ Rept.\  {\bf 513}, 1 (2012)}
  [\href{\arxiv/arXiv:1106.2476}{arXiv:1106.2476} [astro-ph.CO]].
  
\bibitem{Hinterbichler:2011tt} 
  K.~Hinterbichler,
  \href{\doi/10.1103/RevModPhys.84.671}{Rev.\ Mod.\ Phys.\  {\bf 84}, 671 (2012)}
  [\href{\arxiv/arXiv:1105.3735}{arXiv:1105.3735} [hep-th]].
  
\bibitem{Dvali:2000hr} 
  G.~R.~Dvali, G.~Gabadadze and M.~Porrati,
  \href{\doi/10.1016/S0370-2693(00)00669-9}{Phys.\ Lett.\ B {\bf 485}, 208 (2000)}
  [\href{\arxiv/hep-th/0005016}{hep-th/0005016}].
  
\bibitem{Nicolis:2008in} 
  A.~Nicolis, R.~Rattazzi and E.~Trincherini,
  \href{\doi/10.1103/PhysRevD.79.064036}{Phys.\ Rev.\ D {\bf 79}, 064036 (2009)}
  [\href{\arxiv/arXiv:0811.2197}{arXiv:0811.2197} [hep-th]].
  
  
\bibitem{deRham:2012az} 
  C.~de Rham,
  \href{\doi/10.1016/j.crhy.2012.04.006}{Comptes Rendus Physique {\bf 13}, 666 (2012)}
  [\href{\arxiv/arXiv:1204.5492}{arXiv:1204.5492} [astro-ph.CO]].
  
\bibitem{deRham:2010kj} 
  C.~de Rham, G.~Gabadadze and A.~J.~Tolley,
  \href{\doi/10.1103/PhysRevLett.106.231101}{Phys.\ Rev.\ Lett.\  {\bf 106}, 231101 (2011)}
  [\href{\arxiv/arXiv:1011.1232}{arXiv:1011.1232} [hep-th]].
  
\bibitem{deRham:2014zqa} 
  C.~de Rham,
  \href{\doi/10.12942/lrr-2014-7}{Living Rev.\ Rel.\  {\bf 17}, 7 (2014)}
  [\href{\arxiv/arXiv:1401.4173}{arXiv:1401.4173} [hep-th]].
  
\bibitem{DeFelice:2010aj} 
  A.~De Felice and S.~Tsujikawa,
  \href{\doi/10.12942/lrr-2010-3}{Living Rev.\ Rel.\  {\bf 13}, 3 (2010)}
  [\href{\arxiv/arXiv:1002.4928}{arXiv:1002.4928} [gr-qc]].
  


\bibitem{Khoury:2003aq} 
  J.~Khoury and A.~Weltman,
  \href{\doi/10.1103/PhysRevLett.93.171104}{Phys.\ Rev.\ Lett.\  {\bf 93}, 171104 (2004)}
  [\href{\arxiv/astro-ph/0309300}{astro-ph/0309300}].
  
\bibitem{Khoury:2003rn} 
  J.~Khoury and A.~Weltman,
  \href{\doi/10.1103/PhysRevD.69.044026}{Phys.\ Rev.\ D {\bf 69}, 044026 (2004)}
  [\href{\arxiv/astro-ph/0309411}{astro-ph/0309411}].
  
    
\bibitem{Hinterbichler:2010es} 
  K.~Hinterbichler and J.~Khoury,
  \href{\doi/10.1103/PhysRevLett.104.231301}{Phys.\ Rev.\ Lett.\  {\bf 104}, 231301 (2010)}
  [\href{\arxiv/arXiv:1001.4525}{arXiv:1001.4525} [hep-th]].
  
\bibitem{Vainshtein:1972sx} 
  A.~I.~Vainshtein,
  \href{\doi/10.1016/0370-2693(72)90147-5}{Phys.\ Lett.\  {\bf 39B}, 393 (1972)}.
  
\bibitem{vanDam:1970vg} 
  H.~van Dam and M.~J.~G.~Veltman,
  \href{\doi/10.1016/0550-3213(70)90416-5}{Nucl.\ Phys.\ B {\bf 22}, 397 (1970)}.
  
\bibitem{Zakharov:1970cc} 
  V.~I.~Zakharov,
  JETP Lett.\  {\bf 12}, 312 (1970)
  [Pisma Zh.\ Eksp.\ Teor.\ Fiz.\  {\bf 12}, 447 (1970)].
  
\bibitem{Fierz:1939ix} 
  M.~Fierz and W.~Pauli,
  \href{\doi/10.1098/rspa.1939.0140}{Proc.\ Roy.\ Soc.\ Lond.\ A {\bf 173}, 211 (1939)}.
  


\bibitem{Luty:2003vm} 
  M.~A.~Luty, M.~Porrati and R.~Rattazzi,
  \href{\doi/10.1088/1126-6708/2003/09/029}{JHEP {\bf 0309}, 029 (2003)}
  [\href{\arxiv/hep-th/0303116}{hep-th/0303116}].

\bibitem{Nicolis:2004qq} 
  A.~Nicolis and R.~Rattazzi,
  \href{\doi/10.1088/1126-6708/2004/06/059}{JHEP {\bf 0406}, 059 (2004)}
  [\href{\arxiv/hep-th/0404159}{hep-th/0404159}].
  
\bibitem{Woodard:2006nt} 
  R.~P.~Woodard,
  \href{\doi/10.1007/978-3-540-71013-4_14}{Lect.\ Notes Phys.\  {\bf 720}, 403 (2007)}
  [\href{\arxiv/astro-ph/0601672}{astro-ph/0601672}].
  
\bibitem{Deffayet:2009wt} 
  C.~Deffayet, G.~Esposito-Farese and A.~Vikman,
  \href{\doi/10.1103/PhysRevD.79.084003}{Phys.\ Rev.\ D {\bf 79}, 084003 (2009)}
  [\href{\arxiv/arXiv:0901.1314}{arXiv:0901.1314} [hep-th]].
  
\bibitem{Chow:2009fm} 
  N.~Chow and J.~Khoury,
  \href{\doi/10.1103/PhysRevD.80.024037}{Phys.\ Rev.\ D {\bf 80}, 024037 (2009)}
  [\href{\arxiv/arXiv:0905.1325}{arXiv:0905.1325} [hep-th]].
  
\bibitem{Silva:2009km} 
  F.~P.~Silva and K.~Koyama,
  \href{\doi/10.1103/PhysRevD.80.121301}{Phys.\ Rev.\ D {\bf 80}, 121301 (2009)}
  [\href{\arxiv/arXiv:0909.4538}{arXiv:0909.4538} [astro-ph.CO]].
  
\bibitem{Kobayashi:2010wa} 
  T.~Kobayashi,
  \href{\doi/10.1103/PhysRevD.81.103533}{Phys.\ Rev.\ D {\bf 81}, 103533 (2010)}
  [\href{\arxiv/arXiv:1003.3281}{arXiv:1003.3281} [astro-ph.CO]].
  
\bibitem{Kobayashi:2009wr} 
  T.~Kobayashi, H.~Tashiro and D.~Suzuki,
  \href{\doi/10.1103/PhysRevD.81.063513}{Phys.\ Rev.\ D {\bf 81}, 063513 (2010)}
  [\href{\arxiv/arXiv:0912.4641}{arXiv:0912.4641} [astro-ph.CO]].
  
\bibitem{Gannouji:2010au} 
  R.~Gannouji and M.~Sami,
  \href{\doi/10.1103/PhysRevD.82.024011}{Phys.\ Rev.\ D {\bf 82}, 024011 (2010)}
  [\href{\arxiv/arXiv:1004.2808}{arXiv:1004.2808} [gr-qc]].
  
\bibitem{DeFelice:2010gb} 
  A.~De Felice, S.~Mukohyama and S.~Tsujikawa,
  \href{\doi/10.1103/PhysRevD.82.023524}{Phys.\ Rev.\ D {\bf 82}, 023524 (2010)}
  [\href{\arxiv/arXiv:1006.0281}{arXiv:1006.0281} [astro-ph.CO]].
  
\bibitem{DeFelice:2010pv} 
  A.~De Felice and S.~Tsujikawa,
  \href{\doi/10.1103/PhysRevLett.105.111301}{Phys.\ Rev.\ Lett.\  {\bf 105}, 111301 (2010)}
  [\href{\arxiv/arXiv:1007.2700}{arXiv:1007.2700} [astro-ph.CO]].
  
\bibitem{Ali:2010gr} 
  A.~Ali, R.~Gannouji and M.~Sami,
  \href{\doi/10.1103/PhysRevD.82.103015}{Phys.\ Rev.\ D {\bf 82}, 103015 (2010)}
  [\href{\arxiv/arXiv:1008.1588}{arXiv:1008.1588} [astro-ph.CO]].
  
\bibitem{Mota:2010bs} 
  D.~F.~Mota, M.~Sandstad and T.~Zlosnik,
  \href{\doi/10.1007/JHEP12(2010)051}{JHEP {\bf 1012}, 051 (2010)}
  [\href{\arxiv/arXiv:1009.6151}{arXiv:1009.6151} [astro-ph.CO]].
  
\bibitem{Deffayet:2010qz} 
  C.~Deffayet, O.~Pujolas, I.~Sawicki and A.~Vikman,
  \href{\doi/10.1088/1475-7516/2010/10/026}{JCAP {\bf 1010}, 026 (2010)}
  [\href{\arxiv/arXiv:1008.0048}{arXiv:1008.0048} [hep-th]].
  
\bibitem{deRham:2010tw} 
  C.~de Rham, G.~Gabadadze, L.~Heisenberg and D.~Pirtskhalava,
  \href{\doi/10.1103/PhysRevD.83.103516}{Phys.\ Rev.\ D {\bf 83}, 103516 (2011)}
  [\href{\arxiv/arXiv:1010.1780}{arXiv:1010.1780} [hep-th]].
  
\bibitem{deRham:2011by} 
  C.~de Rham and L.~Heisenberg,
  \href{\doi/10.1103/PhysRevD.84.043503}{Phys.\ Rev.\ D {\bf 84}, 043503 (2011)}
  [\href{\arxiv/arXiv:1106.3312}{arXiv:1106.3312} [hep-th]].
  
\bibitem{Heisenberg:2014kea} 
  L.~Heisenberg, R.~Kimura and K.~Yamamoto,
  \href{\doi/10.1103/PhysRevD.89.103008}{Phys.\ Rev.\ D {\bf 89}, 103008 (2014)}
  [\href{\arxiv/arXiv:1403.2049}{arXiv:1403.2049} [hep-th]].
  
\bibitem{Creminelli:2010ba} 
  P.~Creminelli, A.~Nicolis and E.~Trincherini,
  \href{\doi/10.1088/1475-7516/2010/11/021}{JCAP {\bf 1011}, 021 (2010)}
  [\href{\arxiv/arXiv:1007.0027}{arXiv:1007.0027} [hep-th]].
  
\bibitem{Kobayashi:2010cm} 
  T.~Kobayashi, M.~Yamaguchi and J.~Yokoyama,
  \href{\doi/10.1103/PhysRevLett.105.231302}{Phys.\ Rev.\ Lett.\  {\bf 105}, 231302 (2010)}
  [\href{\arxiv/arXiv:1008.0603}{arXiv:1008.0603} [hep-th]].
  
\bibitem{Mizuno:2010ag} 
  S.~Mizuno and K.~Koyama,
  \href{\doi/10.1103/PhysRevD.82.103518}{Phys.\ Rev.\ D {\bf 82}, 103518 (2010)}
  [\href{\arxiv/arXiv:1009.0677}{arXiv:1009.0677} [hep-th]].
  
\bibitem{Burrage:2010cu} 
  C.~Burrage, C.~de Rham, D.~Seery and A.~J.~Tolley,
  \href{\doi/10.1088/1475-7516/2011/01/014}{JCAP {\bf 1101}, 014 (2011)}
  [\href{\arxiv/arXiv:1009.2497}{arXiv:1009.2497} [hep-th]].
  
\bibitem{Kamada:2010qe} 
  K.~Kamada, T.~Kobayashi, M.~Yamaguchi and J.~Yokoyama,
  \href{\doi/10.1103/PhysRevD.83.083515}{Phys.\ Rev.\ D {\bf 83}, 083515 (2011)}
  [\href{\arxiv/arXiv:1012.4238}{arXiv:1012.4238} [astro-ph.CO]].
  
  
\bibitem{Horndeski:1974wa} 
  G.~W.~Horndeski,
  \href{\doi/10.1007/BF01807638}{Int.\ J.\ Theor.\ Phys.\  {\bf 10}, 363 (1974).}
  
\bibitem{Ali:2012cv} 
  A.~Ali, R.~Gannouji, M.~W.~Hossain and M.~Sami,
  \href{\doi/10.1016/j.physletb.2012.10.009}{Phys.}
  \href{\doi/10.1016/j.physletb.2012.10.009}{\ Lett.\ B {\bf 718}, 5 (2012)}
  [\href{\arxiv/arXiv:1207.3959}{arXiv:1207.3959} [gr-qc]].
  
\bibitem{Hossain:2012qm} 
  M.~W.~Hossain and A.~A.~Sen,
  \href{\doi/10.1016/j.physletb.2012.06.016}{Phys.\ Lett.\ B {\bf 713}, 140}
  \href{\doi/10.1016/j.physletb.2012.06.016}{(2012)}
  [\href{\arxiv/arXiv:1201.6192}{arXiv:1201.6192} [astro-ph.CO]].
  
  
\bibitem{Bartolo:2013ws} 
  N.~Bartolo, E.~Bellini, D.~Bertacca and S.~Matarrese,
  \href{\doi/10.1088/1475-7516/2013/03/034}{JCAP {\bf 1303}, 034 (2013)}
  [\href{\arxiv/arXiv:1301.4831}{arXiv:1301.4831} [astro-ph.CO]].
  
\bibitem{Koyama:2009me} 
  K.~Koyama, A.~Taruya and T.~Hiramatsu,
  \href{\doi/10.1103/PhysRevD.79.123512}{Phys.\ Rev.\ D {\bf 79}, 123512 (2009)}
  [\href{\arxiv/arXiv:0902.0618}{arXiv:0902.0618} [astro-ph.CO]].
  
\bibitem{Kimura:2010di} 
  R.~Kimura and K.~Yamamoto,
  \href{\doi/10.1088/1475-7516/2011/04/025}{JCAP {\bf 1104}, 025 (2011)}
  [\href{\arxiv/arXiv:1011.2006}{arXiv:1011.2006} [astro-ph.CO]].
  
\bibitem{Brax:2012sy} 
  P.~Brax and P.~Valageas,
  \href{\doi/10.1103/PhysRevD.86.063512}{Phys.\ Rev.\ D {\bf 86}, 063512 (2012)}
  [\href{\arxiv/arXiv:1205.6583}{arXiv:1205.6583} [astro-ph.CO]].
  
\bibitem{Barreira:2013eea} 
  A.~Barreira, B.~Li, W.~A.~Hellwing, C.~M.~Baugh and S.~Pascoli,
  \href{\doi/10.1088/1475-7516/2013/10/027}{JCAP {\bf 1310}, 027 (2013)}
  [\href{\arxiv/arXiv:1306.3219}{arXiv:1306.3219} [astro-ph.CO]].
  
\bibitem{Li:2013tda} 
  B.~Li, A.~Barreira, C.~M.~Baugh, W.~A.~Hellwing, K.~Koyama, S.~Pascoli and G.~B.~Zhao,
  \href{\doi/10.1088/1475-7516/2013/11/012}{JCAP {\bf 1311}, 012 (2013)}
  [\href{\arxiv/arXiv:1308.3491}{arXiv:1308.3491} [astro-ph.CO]].
  
\bibitem{Wyman:2013jaa} 
  M.~Wyman, E.~Jennings and M.~Lima,
  \href{\doi/10.1103/PhysRevD.88.084029}{Phys.\ Rev.\ D {\bf 88}, no. 8, 084029 (2013)}
  doi:10.1103/PhysRevD.88.084029
  [\href{\arxiv/arXiv:1303.6630}{arXiv:1303.6630} [astro-ph.CO]].
  
\bibitem{Takushima:2013foa} 
  Y.~Takushima, A.~Terukina and K.~Yamamoto,
  \href{\doi/10.1103/PhysRevD.89.104007}{Phys.\ Rev.\ D {\bf 89}, no. 10, 104007 (2014)}
  [\href{\arxiv/arXiv:1311.0281}{arXiv:1311.0281} [astro-ph.CO]].
  
\bibitem{Taruya:2014faa} 
  A.~Taruya, T.~Nishimichi, F.~Bernardeau, T.~Hiramatsu and K.~Koyama,
  \href{\doi/10.1103/PhysRevD.90.123515}{Phys.\ Rev.\ D {\bf 90}, no. 12, 123515 (2014)}
  [\href{\arxiv/arXiv:1408.4232}{arXiv:1408.4232} [astro-ph.CO]].
  
\bibitem{Takushima:2015iha} 
  Y.~Takushima, A.~Terukina and K.~Yamamoto,
  \href{\doi/10.1103/PhysRevD.92.104033}{Phys.\ Rev.\ D {\bf 92}, no. 10, 104033 (2015)}
  [\href{\arxiv/arXiv:1502.03935}{arXiv:1502.03935} [gr-qc]].
  
\bibitem{Bellini:2015wfa} 
  E.~Bellini, R.~Jimenez and L.~Verde,
  \href{\doi/10.1088/1475-7516/2015/05/057}{JCAP {\bf 1505}, no. 05, 057 (2015)}
  [arXiv:1504.04341 [astro-ph.CO]].
  
  
  
\bibitem{Hudson:2012zga} 
  M.~J.~Hudson, S.~J.~Turnbull,
  \href{\doi/10.1088/2041-8205/751/2/L30}{Astrophys.\ J.\ Lett.\ {\bf 751}, L30 (2012)}
  [\href{\arxiv/arXiv:1203.4814}{arXiv:1203.4814} [astro-ph.CO]].
  
\bibitem{Beutler:2012px} 
  F.~Beutler {\it et al.},
  \href{\doi/10.1111/j.1365-2966.2012.21136.x}{Mon.\ Not.\ Roy.\ Astron.\ Soc.\  {\bf 423}, 3430 (2012)}
  [\href{\arxiv/arXiv:1204.4725}{arXiv:1204.4725} [astro-ph.CO]].
  
\bibitem{Howlett:2014opa} 
  C.~Howlett, A.~Ross, L.~Samushia, W.~Percival and M.~Manera,
  \href{\doi/10.1093/mnras/stu2693}{Mon.\ Not.\ Roy.\ Astron.\ Soc.\  {\bf 449}, no. 1, 848 (2015)}
  [\href{\arxiv/arXiv:1409.3238}{arXiv:1409.3238} [astro-ph.CO]].
  
\bibitem{Percival:2004fs} 
  W.~J.~Percival {\it et al.} [2dFGRS Collaboration],
  \href{\doi/10.1111/j.1365-2966.2004.08146.x}{Mon.\ Not.\ Roy.\ Astron.\ Soc.\  {\bf 353}, 1201 (2004)}
  [\href{\arxiv/astro-ph/0406513}{astro-ph/0406513}].
  
\bibitem{Song:2008qt} 
  Y.~S.~Song and W.~J.~Percival,
  \href{\doi/10.1088/1475-7516/2009/10/004}{JCAP {\bf 0910}, 004 (2009)}
  [\href{\arxiv/arXiv:0807.0810}{arXiv:0807.0810} [astro-ph]].
  
\bibitem{Blake:2011rj} 
  C.~Blake {\it et al.},
  \href{\doi/10.1111/j.1365-2966.2011.18903.x}{Mon.\ Not.\ Roy.\ Astron.\ Soc.\  {\bf 415}, 2876 (2011)}
  [\href{\arxiv/arXiv:1104.2948}{arXiv:1104.2948} [astro-ph.CO]].
  
\bibitem{Samushia:2011cs} 
  L.~Samushia, W.~J.~Percival and A.~Raccanelli,
  \href{\doi/10.1111/j.1365-2966.2011.20169.x}{Mon.\ Not.\ Roy.\ Astron.\ Soc.\  {\bf 420}, 2102 (2012)}
  [\href{\arxiv/arXiv:1102.1014}{arXiv:1102.1014} [astro-ph.CO]].
  
\bibitem{Tojeiro:2012rp} 
  R.~Tojeiro {\it et al.},
  \href{\doi/10.1111/j.1365-2966.2012.21404.x}{Mon.\ Not.\ Roy.\ Astron.\ Soc.\  {\bf 424}, 2339 (2012)}
  [\href{\arxiv/arXiv:1203.6565}{arXiv:1203.6565} [astro-ph.CO]].
  
\bibitem{Gil-Marin:2015sqa} 
  H.~Gil-Marín {\it et al.},
  \href{\doi/10.1093/mnras/stw1096}{Mon.\ Not.\ Roy.\ Astron.\ Soc.\  {\bf 460}, no. 4, 4188 (2016)}
  [\href{\arxiv/arXiv:1509.06386}{arXiv:1509.06386} [astro-ph.CO]].
  
\bibitem{Tegmark:2006az} 
  M.~Tegmark {\it et al.} [SDSS Collaboration],
  \href{\doi/10.1103/PhysRevD.74.123507}{Phys.\ Rev.\ D {\bf 74}, 123507 (2006)}
  [\href{\arxiv/astro-ph/0608632}{astro-ph/0608632}].
  
\bibitem{Blake:2012pj} 
  C.~Blake {\it et al.},
  \href{\doi/10.1111/j.1365-2966.2012.21473.x}{Mon.\ Not.\ Roy.\ Astron.\ Soc.\  {\bf 425}, 405 (2012)}
  [\href{\arxiv/arXiv:1204.3674}{arXiv:1204.3674} [astro-ph.CO]].
  
\bibitem{Chuang:2013wga} 
  C.~H.~Chuang {\it et al.},
  \href{\doi/10.1093/mnras/stw1535}{Mon.\ Not.\ Roy.\ Astron.\ Soc.\  {\bf 461}, no. 4, 3781 (2016)}
  [\href{\arxiv/arXiv:1312.4889}{arXiv:1312.4889} [astro-ph.CO]].
  
\bibitem{Macaulay:2013swa} 
  E.~Macaulay, I.~K.~Wehus and H.~K.~Eriksen,
  \href{\doi/10.1103/PhysRevLett.111.161301}{Phys.\ Rev.\ Lett.\  {\bf 111}, no. 16, 161301 (2013)}
  [\href{\arxiv/arXiv:1303.6583}{arXiv:1303.6583} [astro-ph.CO]].
  
\bibitem{Guzzo:2008ac} 
  L.~Guzzo {\it et al.},
  \href{\doi/10.1038/nature06555}{Nature {\bf 451}, 541 (2008)}
  [\href{\arxiv/arXiv:0802.1944}{arXiv:0802.1944} [astro-ph]].
  
\bibitem{delaTorre:2013rpa} 
  S.~de la Torre {\it et al.},
  \href{\doi/10.1051/0004-6361/201321463}{Astron.\ Astrophys.\  {\bf 557}, A54 (2013)}
  [\href{\arxiv/arXiv:1303.2622}{arXiv:1303.2622} [astro-ph.CO]].
  
\bibitem{Okumura:2015lvp} 
  T.~Okumura {\it et al.},
  \href{\doi/10.1093/pasj/psw029}{Publ.\ Astron.\ Soc.\ Jap.\  {\bf 68}, no. 3, id. 38, 24 (2016)}
  [\href{\arxiv/arXiv:1511.08083}{arXiv:1511.08083} [astro-ph.CO]].
  

  
  

\bibitem{Okada:2012mn} 
  H.~Okada, T.~Totani and S.~Tsujikawa,
  \href{\doi/10.1103/PhysRevD.87.103002}{Phys.\ Rev.\ D {\bf 87}, no. 10, 103002 (2013)}
  [\href{\arxiv/arXiv:1208.4681}{arXiv:1208.4681} [astro-ph.CO]].
  
\bibitem{Tsujikawa:2012hv} 
  S.~Tsujikawa, A.~De Felice and J.~Alcaniz,
  \href{\doi/10.1088/1475-7516/2013/01/030}{JCAP {\bf 1301}, 030 (2013)}
  [\href{\arxiv/arXiv:1210.4239}{arXiv:1210.4239} [astro-ph.CO]].

\bibitem{Dodelson:2013sma} 
  S.~Dodelson and S.~Park,
  \href{\doi/10.1103/PhysRevD.90.043535}{Phys.\ Rev.\ D {\bf 90}, 043535 (2014)}
  [\href{\arxiv/arXiv:1310.4329}{arXiv:1310.4329} [astro-ph.CO]].
  
\bibitem{Nesseris:2014mea} 
  S.~Nesseris and S.~Tsujikawa,
  \href{\doi/10.1103/PhysRevD.90.024070}{Phys.\ Rev.\ D {\bf 90}, no. 2, 024070 (2014)}
  [\href{\arxiv/arXiv:1402.4613}{arXiv:1402.4613} [astro-ph.CO]].
  
\bibitem{DeFelice:2016ufg} 
  A.~De Felice and S.~Mukohyama,
  \href{\doi/10.1103/PhysRevLett.118.091104}{Phys.\ Rev.\ Lett.\  {\bf 118}, no. 9, 091104 (2017)}
  [\href{\arxiv/arXiv:1607.03368}{arXiv:1607.03368} [astro-ph.CO]].
  
\bibitem{Park:2016jym} 
  S.~Park and A.~Shafieloo,
  \href{\doi/10.1103/PhysRevD.95.064061}{Phys.\ Rev.\ D {\bf 95}, no. 6, 064061 (2017)}
  [\href{\arxiv/arXiv:1608.02541}{arXiv:1608.02541} [astro-ph.CO]].
  
\bibitem{Nersisyan:2017mgj} 
  H.~Nersisyan, A.~F.~Cid and L.~Amendola,
  \href{\arxiv/arXiv:1701.00434}{arXiv:1701.00434} [astro-ph.CO].
  
  
\bibitem{Ade:2015lrj} 
  P.~A.~R.~Ade {\it et al.} [Planck Collaboration],
  \href{\doi/10.1051/0004-6361/201525898}{Astron.\ Astrophys.\  {\bf 594}, A20 (2016)}
  [\href{\arxiv/arXiv:1502.02114}{arXiv:1502.02114} [astro-ph.CO]].
  
\bibitem{Bernardeau:2001qr} 
  F.~Bernardeau, S.~Colombi, E.~Gaztanaga and R.~Scoccimarro,
  \href{\doi/10.1016/S0370-1573(02)00135-7}{Phys.\ Rept.\  {\bf 367}, 1 (2002)}
  [\href{\arxiv/astro-ph/0112551}{astro-ph/0112551}].




\bibitem{Deruelle:2010ht} 
  N.~Deruelle and M.~Sasaki,
  \href{\doi/10.1007/978-3-642-19760-4_23}{Springer Proc.\ Phys.\  {\bf 137}, 247 (2011)}
  [\href{\arxiv/arXiv:1007.3563}{arXiv:1007.3563} [gr-qc]].
  
\bibitem{Gong:2011qe} 
  J.~O.~Gong, J.~c.~Hwang, W.~I.~Park, M.~Sasaki and Y.~S.~Song,
  \href{\doi/10.1088/1475-7516/2011/09/023}{JCAP {\bf 1109}, 023 (2011)}
  [\href{\arxiv/arXiv:1107.1840}{arXiv:1107.1840} [gr-qc]].
  
\bibitem{Chiba:2013mha} 
  T.~Chiba and M.~Yamaguchi,
  \href{\doi/10.1088/1475-7516/2013/10/040}{JCAP {\bf 1310}, 040 (2013)}
  [\href{\arxiv/arXiv:1308.1142}{arXiv:1308.1142} [gr-qc]].




\bibitem{Amendola:1999er} 
  L.~Amendola,
  \href{\doi/10.1103/PhysRevD.62.043511}{Phys.\ Rev.\ D {\bf 62}, 043511 (2000)}
  [\href{\arxiv/astro-ph/9908023}{as}
  \href{\arxiv/astro-ph/9908023}{tro-ph/9908023}].
  

\bibitem{Barreiro:1999zs} 
  T.~Barreiro, E.~J.~Copeland and N.~J.~Nunes,
  \href{\doi/10.1103/PhysRevD.61.127301}{Phys.\ Rev.\ D {\bf 61}, 127301 (2000)}
  [\href{\arxiv/astro-ph/9910214}{astro-ph/9910214}].
  
\bibitem{Sahni:1999qe} 
  V.~Sahni and L.~M.~Wang,
  \href{\doi/10.1103/PhysRevD.62.103517}{Phys.\ Rev.\ D {\bf 62}, 103517 (2000)}
  doi:10.1103/PhysRevD.62.103517
  [astro-ph/9910097].

  
\bibitem{Bardeen:1980kt} 
  J.~M.~Bardeen,
  \href{\doi/10.1103/PhysRevD.22.1882}{Phys.\ Rev.\ D {\bf 22}, 1882 (1980)}.
  
\bibitem{Eisenstein:1997jh} 
  D.~J.~Eisenstein and W.~Hu,
  \href{\doi/10.1086/306640}{Astrophys.\ J.\  {\bf 511}, 5 (1997)}
  [\href{\arxiv/astro-ph/9710252}{astro-ph/9710252}].
  
\bibitem{Duniya:2015nva} 
  D.~G.~A.~Duniya, D.~Bertacca and R.~Maartens,
  \href{\doi/10.1103/PhysRevD.91.063530}{Phys.\ Rev.}
  \href{\doi/10.1103/PhysRevD.91.063530}{D {\bf 91}, 063530 (2015)}
  [\href{\arxiv/arXiv:1502.06424}{arXiv:1502.06424} [astro-ph.CO]].
  
\bibitem{Eisenstein:1997ik} 
  D.~J.~Eisenstein and W.~Hu,
  \href{\doi/10.1086/305424}{Astrophys.\ J.\  {\bf 496}, 605 (1998)}
  [\href{\arxiv/astro-ph/9709112}{astro-ph/9709112}].
  


\end{thebibliography}
\end{document}